\PassOptionsToPackage{hyphens}{url}
\documentclass{aa}

\usepackage{graphicx}
\usepackage{txfonts}
\usepackage{color}
\usepackage{xspace}
\usepackage{siunitx}
\usepackage{amsmath, amsfonts}
\usepackage{amssymb}
\usepackage{xcolor}
\usepackage{dashbox}
\usepackage{framed}
\usepackage{lipsum}
\usepackage{placeins}
\usepackage{stfloats}
\usepackage{picture}
\usepackage{multirow}
\usepackage{balance}

\usepackage{tikz}
\usepackage[dvipsnames]{xcolor}
\usepackage[normalem]{ulem}

\usetikzlibrary{shapes.geometric, arrows, shapes.misc}
\usetikzlibrary{calc}

\tikzstyle{startstop} = [rectangle, rounded corners, minimum width=3cm, minimum height=0.75cm,text centered, draw=black, fill=Mulberry!40, draw=Mulberry!60,  line width=0.5mm]  
\tikzstyle{process} = [rectangle, rounded corners, minimum width=4.cm, minimum height=0.75cm,text centered, draw=black, fill=SkyBlue!40, draw=SkyBlue!60,  line width=0.5mm] 
\tikzstyle{process_small} = [rectangle, rounded corners, minimum width=0cm, minimum height=0.75cm,text centered, draw=black, fill=SkyBlue!40, draw=SkyBlue!60,  line width=0.5mm] 
\tikzstyle{process_big} = [rectangle, rounded corners, minimum width=9cm, minimum height=0.75cm,text centered, draw=black, fill=SkyBlue!40, draw=SkyBlue!60,  line width=0.5mm] 
\tikzstyle{decision} = [rectangle,  minimum width=2cm, minimum height=0.5cm,text centered, fill=Yellow!40, draw=black]
\tikzstyle{product} = [rectangle, rounded corners, minimum width=3.25cm, minimum height=0.75cm,text centered, fill=Dandelion!40, draw=Dandelion!60, line width=0.5mm]
\tikzstyle{product1} = [rectangle, rounded corners, minimum width=2cm, minimum height=0.75cm,text centered, fill=Gray!40, draw=Gray!60, line width=0.5mm]
\tikzstyle{arrow} = [thick,->,>=stealth]
\tikzstyle{arrow2} = [thick,-,>=stealth]
\tikzstyle{branch} = [chamfered rectangle , chamfered rectangle xsep=2cm, minimum width=4.cm, minimum height=0.75cm,text centered, fill=SeaGreen!40, draw=SeaGreen!60, line width=0.5mm]

\tikzstyle{startstop2} = [rectangle, rounded corners, minimum width=3cm, minimum height=0.6cm,text centered, draw=black, fill=SeaGreen!40, draw=SeaGreen!60,  line width=0.5mm]  
\tikzstyle{process2} = [rectangle, rounded corners, minimum width=4.5cm, minimum height=0.6cm,text centered, draw=black, fill=SkyBlue!40, draw=SkyBlue!60,  line width=0.5mm] 
\tikzstyle{process_small2} = [rectangle, rounded corners, minimum width=0cm, minimum height=0.6cm,text centered, draw=black, fill=SkyBlue!40, draw=SkyBlue!60,  line width=0.5mm] 
\tikzstyle{process_big2} = [rectangle, rounded corners, minimum width=8.5cm, minimum height=0.6cm,text centered, draw=black, fill=SkyBlue!40, draw=SkyBlue!60,  line width=0.5mm] 
\tikzstyle{decision2} = [rectangle,  minimum width=2cm, minimum height=0.6cm,text centered, fill=Yellow!40, draw=black]
\tikzstyle{product2} = [rectangle, rounded corners, minimum width=2cm, minimum height=0.6cm,text centered, fill=Dandelion!40, draw=Dandelion!60, line width=0.5mm]
\tikzstyle{product_unimportant2} = [rectangle, rounded corners, minimum width=3cm, minimum height=0.6cm,text centered, fill=Gray!40, draw=Gray!60, line width=0.5mm]
\tikzstyle{process_unimportant2} = [rectangle, rounded corners, minimum width=4.5cm, minimum height=0.6cm,text centered, fill=Gray!40, draw=Gray!60, line width=0.5mm]
\tikzstyle{product_important2} = [rectangle, rounded corners, minimum width=3cm, minimum height=0.6cm,text centered, fill=Mulberry!40, draw=Mulberry!60, line width=0.5mm]
\tikzstyle{arrow} = [thick,->,>=stealth]

\tikzstyle{core} = [rectangle, minimum width=6.2cm, minimum height=6cm, text centered, text width=5cm, draw=black, fill=orange!30]
\tikzstyle{epochs} = [rectangle, minimum width=7cm, minimum height=6cm, text centered, text width=6cm, draw=black, fill=SeaGreen!30]
\tikzstyle{crossmatch} = [rectangle, minimum width=6cm, minimum height=6cm, text centered, text width=4.5cm, draw=black, fill=Mulberry!30]

\usepackage{hyperref}
\hypersetup{
        colorlinks=true,
        breaklinks=true,
        linkcolor=blue,
        citecolor=blue,
        filecolor=blue,
        urlcolor=blue,
        unicode=false,
        pdftoolbar=true,
        pdfmenubar=true,
        pdffitwindow=false,
        pdfstartview={Fit},
        pdftitle={},
        pdfauthor={Alena Rottensteiner},
        pdfsubject={},
        pdfcreator={Alena Rottensteiner},
        pdfkeywords={},
        pdfnewwindow=true,
        pdfdisplaydoctitle=true
}

\makeatletter
\renewcommand*\aa@pageof{, page \thepage{} of \pageref*{LastPage}}
\makeatother

\begin{document}

\clearpage
\newpage

\title{Kinematics of young stellar objects in NGC~2024 based on infrared proper motions}

\author{Alena Rottensteiner\inst{1,2}   \and 
        Monika G. Petr-Gotzens\inst{2}  \and 
        Stefan Meingast\inst{1}         \and
        João Alves\inst{1,3}            \and
        Emmanuel Bertin\inst{4}         \and 
        Hervé Bouy\inst{5,6}            \and
        Martin Piecka\inst{1}           \and
        Sebastian Ratzenböck\inst{7}    \and
        Andrea Socci\inst{1}
        }
              
\institute{Department of Astrophysics, University of Vienna, T\"urkenschanzstrasse 17, 1180 Wien, Austria 
\\ \email{alena.kristina.rottensteiner@univie.ac.at}
\and 
European Southern Observatory, Karl-Schwarzschild-Strasse 2, 85748 Garching bei München, Germany
\and
University of Vienna, Research Network Data Science at Uni Vienna, Kolingasse 14-16, 1090 Wien, Austria
\and
Université Paris-Saclay, Université Paris Cité, CEA, CNRS, AIM, 91191 Gif-sur-Yvette, France
\and
Laboratoire d’astrophysique de Bordeaux, Univ. Bordeaux, CNRS, B18N, allée Geoffroy Saint-Hilaire, 33615 Pessac, France
\and
Institut universitaire de France (IUF), 1 rue Descartes, 75231 Paris Cedex 05, France
\and 
Center for Astrophysics | Harvard \& Smithsonian, 60 Garden St., Cambridge, MA 02138, USA
}

\date{Received October 2, 2025 / Accepted November 27, 2025}

\abstract
{
The most recently formed young stellar objects (YSOs) in active star forming regions are excellent tracers of their parent cloud motion. Their positions and dynamics provide insight into cluster formation and constrain kinematic decoupling timescales between stars and gas. However, because of their strong extinction and young age, embedded YSOs are mainly visible at infrared wavelengths and thus absent from astrometric surveys such as \emph{Gaia}.
We measured the proper motions of 6,769 sources toward the NGC~2024 cluster in the Flame Nebula ($d\sim420$ pc) using multi-epoch near-infrared observations from three ESO public surveys: VISIONS, VHS, and the VISTA/VIRCAM science verification program. Cross-validation of our results with \emph{Gaia} using optically visible stars shows excellent agreement, with uncertainties on the same order of magnitude. 
For 362 YSO candidates identified from the literature, we derived proper motions on the order of $<5$ mas\,yr$^{-1}$ with mean measurement uncertainties of ~0.22 mas\,yr$^{-1}$, or 0.44 km\,s$^{-1}$. This is the first homogeneous proper motion measurement of this quality for more than half of these stars. For Class~I and flat-spectrum sources, our results provide a $>$13-fold increase in available proper motion measurements.
We analyzed the positional and kinematic differences between YSO classes and confirmed a previously reported inside-out age segregation from younger to older stars, likely driven by an outward movement of older stars. 
No evidence of prolonged hierarchical assembly was found. Instead, the results support a rapid ($<1$ Myr) cluster collapse into a centrally concentrated system. This scenario also accounts for the observed slightly higher 1D velocity dispersion of Class~I sources relative to Class~flat objects. YSO radial velocities generally align with the gas velocities measured from the molecular transitions of $^{12}$CO$(3-2)$, HNC$(1-0)$, HCN$(1-0)$, and show a weaker correlation with N$_2$H$^+(1-0)$. Some Class~II and III objects appear to be already decoupling.
}


\maketitle

\section{Introduction}
\label{sec:Introduction}

Star formation is one of the most-studied processes in the field of astronomy. Its physical drivers and relative timelines set the initial conditions of planet formation and habitability \citep[e.g.,][]{Dai_2023} and can shape enormous Galactic structures \citep[e.g.,][]{2020Alves, 2024Swiggum}. It is understood that many stars form simultaneously \citep{Lada2003} from a complex network of dense filaments and hubs \citep[e.g.,][]{2009Myers,2010Andre_HGBS,2020Kumar}. As the filaments fragment and collapse to form pre-stellar cores and later young stellar objects (YSOs), they imprint their structural and kinematic properties on the newly formed objects \citep{2016Hacar}. 

However, the formation process of star clusters remains debated. Simulations suggest a hierarchical merging of subclusters at early times, but soon after, their structure can resemble monolithic growth \citep[][and references therein]{2025Laverde}. Recent simulations found that initially hierarchical spatial substructure dissipated within $\leq 2.5$ freefall times, but kinematic signatures persisted longer \citep{2025Laverde}. In contrast, simulations by \cite{2018Sills}, with initial parameters inspired by observations \citep{2019Kuhn}, showed a rapid cluster collapse of $\leq1$ Myr into a monolithic system without any detectable substructure. The collapse scenarios were also shown to change YSO velocity dispersions as a function of their age \citep[e.g.,][]{2009Proskow, 2018Sills}. 

The different formation pathways of clusters demonstrate that simulations strongly depend on the initial conditions and relative timescales of star formation, which are themselves poorly constrained. YSOs are born with a physical memory of their birth cloudlet that — unless the gas still present was recently disturbed by feedback, shocks, or winds — traces the motion of the surrounding material. Over the span of a few $10^5 -10^6$ yr, YSOs decouple kinematically and spatially from the gas. This process is likely partly driven by N-body interactions increasing the stellar velocity dispersion \citep{2003Bate}. At the same time, the gas reservoir gets diminished by both star formation and dispersal by stellar feedback. The timescales of the decoupling remain poorly constrained, making them one of the central open problems in star formation. They are not only essential for understanding the relative roles of stellar feedback, the dynamical state and stability of clusters, and the molecular cloud lifetimes, but also critical for initial conditions of simulations \citep{2025Laverde}. 

At present, observations are the only available method to constrain decoupling timescales and test cluster assembly theories. In this respect, the \emph{Gaia} \citep{Gaia_mission} has dramatically advanced our dynamical understanding of star formation, providing five- or six-dimensional position–velocity data ($\alpha$, $\delta$, $\varpi$, $\mu_{\alpha*}$, $\mu_{\delta}$, $v_{\text{LOS}}$) for over a billion sources \citep{Gaia_DR3}. These data have revealed feedback-driven structures and evolutionary timescales of star-forming complexes that are a few Myr old \citep[][]{2023Ratzenböck, 2023Posch, 2025Posch}. 
However, as an optical instrument, \emph{Gaia} cannot probe the crucial first few Myr of star formation. Stellar nurseries are sites of a phase transition between diffuse gas and compact YSOs, often heavily obscured by dust. Newly formed stars are embedded in high column density regions and surrounded by circumstellar disks that absorb and re-emit radiation at longer wavelengths. Thus, they are optically obscured, but readily detectable from the infrared to the radio regime, depending on their extinction and evolutionary stage. This means that precise and homogeneous kinematic measurements such as those provided by \emph{Gaia} are still lacking for young ($1-5$ Myr) populations with embedded components.

With this project, we propose a complementary method to \emph{Gaia} to address this gap -- accessing the kinematics of embedded YSOs in a young cluster using ground-based multi-epoch infrared observations.

We focus on the NGC~2024 cluster in Orion B \citep[$d \approx 420$ pc,][]{2017Kounkel_dist, 2019Zucker, 2023Cao}, the second most massive active stellar nursery in the Orion molecular cloud complex. It is embedded in the Flame Nebula, an extensive H~\textsc{ii} region extending around a central dark lane with a pronounced molecular ridge in the N-S direction, which absorbs light from the visible to the near infrared range \citep[][and references therein]{2008Meyer}. The age estimates for the cluster generally fall below $0.5-2$ Myr \citep{1996Meyer, 2006Levine, 2014Getman}, with ongoing star formation still suspected in its central regions \citep{2003Skinner}. Evidence suggests a radial age gradient increasing from 0.2 Myr in the core to $\sim$1.5 Myr at a distance of $\sim$1 pc from the center \citep{2014Getman}.
NGC~2024 consists of several hundred stars \citep[e.g.,][$\approx 475$ Class~0/I and II YSOs]{2017Kounkel}, identified primarily via photometric surveys and infrared excess measurements \citep[e.g.,][]{1996Meyer}, or X-rays \citep{2014Getman}. The majority of photometrically detected cluster members appear to be harboring accretion disks \citep[e.g.,][]{2003Skinner, Terwisga2020}. 
Attempts at determining cluster members via positional and kinematic \emph{Gaia} data yielded only a fraction (30-60 stars) of the estimated cluster census \citep[e.g.,][]{2017Kounkel, 2024Zerjal}. Its many embedded members and large kinematic dispersion \citep{2024Zerjal} likely contribute to the cluster's absence in the Unified Cluster Catalog \citep{2023Perren}, which summarizes recent clustering efforts.
 
In this paper, we calculate infrared-based proper motions of the same astrometric quality as \emph{Gaia} for 6,769 sources in the direction of the Flame Nebula, among them 362 YSO candidates. Based on this new kinematic information, we discuss different cluster formation scenarios for NGC~2024 and find the best-fitting one to be a rapid ($<1$ Myr) collapse of any substructure into monolithic conditions. We see an overall good agreement between YSO and gas radial velocities, indicating no significant kinematic decoupling except for some older YSOs.

\section{Data}
\label{sec:Data}

\begin{figure}[t]
    \resizebox{\hsize}{!}{\includegraphics{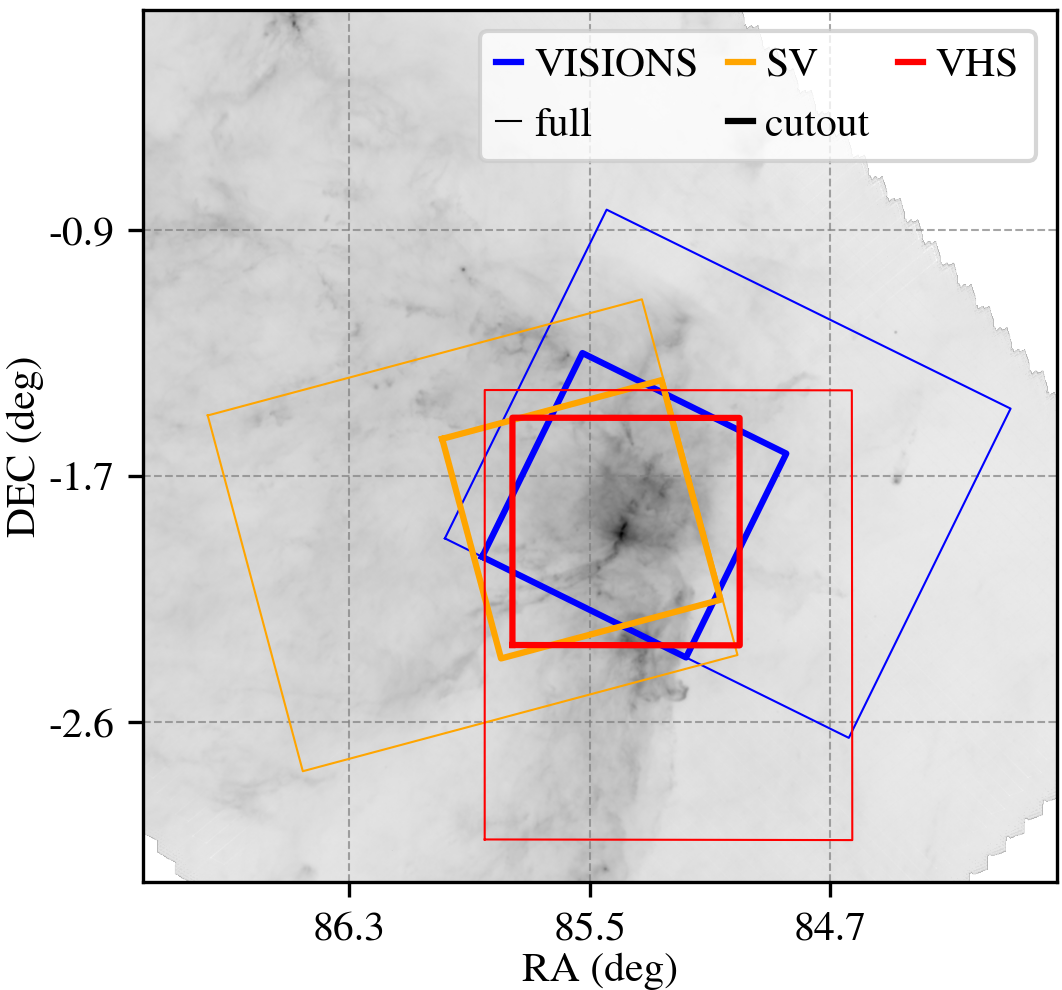}}
    \caption{Outlines of the SV, VHS, and VISIONS tiles covering the Flame Nebula, shown atop the Herschel Gould Belt Survey 350 $\mu$m emission map. The bold squares indicate the cutouts used in this study. The varying tile orientations result from the survey-specific observing strategies.
    }
    \label{fig:01-FN-coverage-tiles}
\end{figure}

Calculating proper motions of embedded stars $d\sim 400$ pc requires observations in infrared passbands with a large temporal baseline and high positional accuracy. For reference, uncertainties in proper motion of $1$ mas\,yr$^{-1}$ correspond to physical uncertainties of around $2$ km\,s$^{-1}$ at this distance. For homogeneity of the results, using the same instrumental setup for all observations is ideal. Wide-field near-infrared surveys are well suited for obtaining spatially complete samples of embedded populations in nearby star forming regions \citep[e.g.,][]{2019Grosschedl}.

We use data from three ESO public surveys carried out with the VISTA/VIRCAM instrument: The Science Verification Galactic mini-survey of Orion \citep[PI: Monika Petr-Gotzens,][ESO program ID 60.A-9285]{2010Arnaboldi, 2011Petr-Gotzens}, hereafter called SV, the Vista Hemisphere Survey \citep[PI: Richard McMahon,][ESO program ID 179.A2010]{2013McMahon}, hereafter called VHS, and the VISIONS survey \citep[PI João Alves,][ESO program ID 198.C-2009]{Meingast2023a}. The SV survey provides one epoch of observations in the $J$, $H$, and $K_s$ bands from October 2009, while VHS contributes an epoch in the $J$ and $K_s$ bands from November 2014. VISIONS adds six $H$-band epochs between 2017 and 2020. Combined, the 11 datasets provide a temporal baseline of $\approx10.3$ years. 

The Flame Nebula is located within a single $\sim1.5^{\circ} \times~1.0^{\circ}$ ``tile'' (see Appendix \ref{appendix:instrumentation-observation}) in each survey. Fig.~\ref{fig:01-FN-coverage-tiles} shows the tile locations atop the Herschel Gould Belt survey 350 $\mu$m map\footnote{\url{http://www.herschel.fr/cea/gouldbelt/en/Phocea/Vie_des_labos/Ast/ast_visu.php?id_ast=66}} of Orion B \citep{2010Andre_HGBS, 2019Arzoumanian_HGBS, 2020Konyves}. For our study, we focused on $0.77^{\circ}~\times~0.77^{\circ}$ cutouts centered on the Flame Nebula, corresponding to an area of around $29$ pc$^2$ at $d\sim420$ pc. The observing parameters for all datasets are listed in Tab.~\ref{tab:02-Observing-parameters}.

\subsection{Science Verification Galactic mini-survey}

The SV galactic-mini survey \citep[][]{2010Arnaboldi, 2011Petr-Gotzens} is a multi-wavelength, high-sensitivity survey of the Orion belt, Orion B, and $\sigma$~Ori populations with a focus on studying the low-mass end of the IMF, searching protostellar disks, and investigating photometric variability in low-mass objects. The observations took place between October 15 and November 2, 2009, and consist of 20 tiles covering an area of about 30 deg$^2$ around the belt stars. 

The tiles were captured with a $15^{\circ}$ tilt in position angle relative to North. The field was observed in the broadband filters $ZYJHK_S$, split into two observation blocks that were carried out in immediate succession for each tile. As the main survey objective was to study low mass objects, its exposure times are longer than those of VISIONS or the VHS, using $8-12$ NDITs but short DIT times of $2-4$ s. Two jitters were performed at each pawprint position. The 5$\sigma$ sensitivity limits of the survey were reported as $Z = 22.5$ mag, $Y = 21.2$ mag, $J = 20.4$ mag, $H = 19.4$ mag, and $K_S = 18.6$ mag \citep{2011Petr-Gotzens}. Further details about the observing strategy are given in \cite{2010Arnaboldi}. The Flame Nebula is located on the survey tile~4.

\subsection{Vista Hemisphere Survey}

The VHS \citep{2013McMahon} is a Cycle 1 program of VISTA\footnote{\url{https://www.eso.org/sci/observing/PublicSurveys/sciencePublicSurveys.html##VISTA}}, which mapped the entire southern hemisphere (Declination $<0^{\circ}$) between 2010 and 2022. The observations were taken in the $J$ and $K_s$ passbands with a reported median $5\sigma$ sensitivity limit of $J = 20.2$ mag and $K_s = 18.1$ mag. The most recent VHS data release, DR5\footnote{\url{https://www.eso.org/qi/catalog/show/290}}, covers around 16,730 square degrees of the southern sky in at least one passband and covers the time frame between November 2009 and March 2017. 

The VHS tiles have a rotator-sky angle of $180^{\circ}$, such that each tile covers about $1.5^{\circ}$ in Right Ascension (RA, $\alpha$) and $1^{\circ}$ in Declination (DEC, $\delta$). They were captured with an effective exposure time $t_{\text{exp}} = 60$s, with long single exposure times and two jitter positions, similar to the SV survey. The Flame Nebula is located on tile 1\_1\_15 in the VHS Stripe 02. The tile was observed during ESO period 94 in November 2014, adding an intermittent data point for proper motion calculations between the SV and VISIONS observations.

\subsection{VISIONS}

 VISIONS \citep{Meingast2023a} is a VISTA Cycle 2 public survey \citep{2019Arnaboldi_Cycle2} designed to produce highly accurate multi-epoch position measurements and proper motions for nearby ($d <$ 500 pc) star forming complexes. It primarily covers low-mass, embedded objects with no \emph{Gaia} coverage. It includes the five nearby regions Chamaeleon, Corona Australis, Lupus, Ophiuchus, and Orion, comprising 650 deg$^2$ of the night sky with roughly 50 h total exposure time. The observations of all regions were taken between April 2017 and March 2022. 

VISIONS consists of a \emph{Wide}, \emph{Deep}, and \emph{Control} field sub-survey. The \emph{Deep} and \emph{Control} sub-surveys cover only small areas and were observed in the $J$, $H$, and $K_S$ filters with long exposure times. Instead, the \emph{Wide} sub-survey was designed to complement the VHS and employed a shallower observing strategy using only the $H$-band over a much larger area. 

Due to its five jitter positions, a VISIONS tile consists of 30 stacked images, whereas a tile comprises 12 images for VHS and SV. The Flame Nebula is on the \emph{Orion wide} tile 1\_9\_6.

\section{Astrometric reduction and proper motion calculation}
\label{sec:Methods}

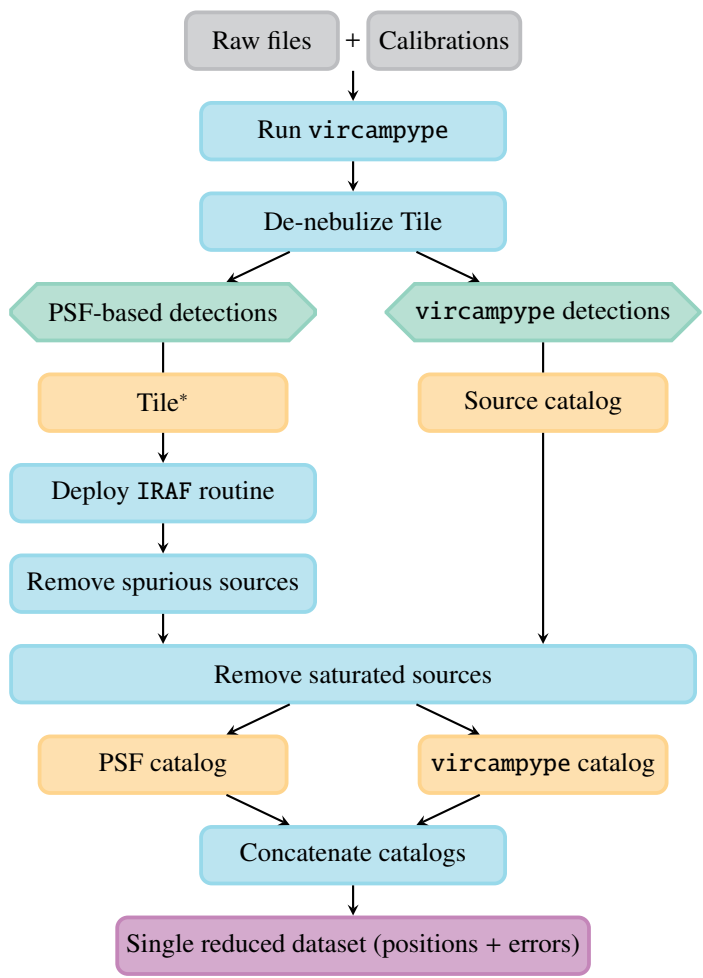
\begin{figure}[t]

\begin{tikzpicture}[node distance=1.2cm]

    \node (plus) {+};
    \node (start1) [product1, left of= plus] {Raw files};
    \node (start2) [product1, right of=plus] {Calibrations};
    \node (vpype) [process, below of=plus] {Run \texttt{vircampype}};
    \node (deneb) [process, below of=vpype] {De-nebulize Tile};
    \node (psf-branch) [branch, below of = deneb, xshift=-2.5cm] {PSF-based detections};
    \node (pype-branch) [branch, below of = deneb, xshift=2.5cm] {\texttt{vircampype} detections};
    \node (tile) [product, below of=psf-branch, ] {Tile$^*$};
    \node (init_cat) [product, below of =pype-branch] {Source catalog};
    \node (IRAF) [process, below of=tile] {Deploy \texttt{IRAF} routine};
    \node (multi) [process, below of=IRAF] {Remove spurious sources};
    \node (sat1) [process_big, below of=multi, xshift=2.5cm] {Remove saturated sources};
    \node (psf) [product, below of=tile, yshift=-3.6cm] {PSF catalog};
    \node (pype) [product, below of=init_cat, yshift=-3.6cm] {\texttt{vircampype} catalog};
    \node (concat) [process, below of=sat1, yshift=-1.2cm] {Concatenate catalogs};
    \node (stop) [startstop, below of=concat] {Single reduced dataset (positions + errors)};

    \draw [arrow] (0,-0.4) -- (plus |- 0, -0.8);
    \draw [arrow] (vpype) -- (deneb) ;
    \draw [arrow] (deneb) --node[anchor=east]{} (psf-branch);
    \draw [arrow] (deneb) --node[anchor=west]{} (pype-branch);
    \draw [arrow2] (psf-branch) -- (tile);
    \draw [arrow2] (pype-branch) -- (init_cat);
    \draw [arrow] (tile) -- (IRAF);
    \draw [arrow] (IRAF) -- (multi);
    \draw [arrow] (multi) -- (multi |- 0, -8.);
    \draw [arrow] (sat1) -- (psf);
    \draw [arrow] (psf) -- (concat);
    \draw [arrow] (init_cat) -- (init_cat |- 0,-8.); %
    \draw [arrow] (sat1) -- (pype);
    \draw [arrow] (pype) -- (concat);
    \draw [arrow] (concat) -- (stop);
    
\end{tikzpicture}

\caption{Astrometric reduction workflow employed on the dataset level.}
\label{fig:02-Workflow-dataset}
\end{figure}

The data processing consists of two workflows, shown in Figs.~\ref{fig:02-Workflow-dataset} and \ref{fig:03-Workflow-master}: First, each of the 11 input files is converted into a table of stellar positions and errors. Secondly, all reduced files are merged into a master catalog, and proper motions are computed.

\subsection{Dataset level}

The science and calibration frames for the VHS and SV tiles, respectively, were downloaded from the ESO Science Archive \footnote{\url{https://archive.eso.org/eso/eso_archive_main.html}}. For VISIONS, we used the raw data from its second data release\footnote{\url{https://www.eso.org/cms/eso-archive-news/second-data-release-from-the-vista-cycle-2-eso-public-survey-visions.html}}. 

\subsubsection{Reduction with \texttt{vircampype}}
The first reduction step on the dataset level was performed with the \texttt{vircampype} data processing pipeline\footnote{\url{https://github.com/smeingast/vircampype}} for VISTA data, which outperformed the Cambridge Astronomical Survey Unit (CASU) pipeline\footnote{\url{http://casu.ast.cam.ac.uk}} on VISIONS test fields \citep{Meingast2023b}. What distinguishes the \texttt{vircampype} procedure is that all input images are calibrated so that the epoch of source coordinates equals the observation date of an image.

The pipeline workflow is shown in \cite{Meingast2023b}, Fig.~1. In short, raw observations and calibrations are transformed into science-ready tiles and stacks, which are used to derive source catalogs containing stellar positions and magnitudes, their statistical errors, and quality flags from source extraction. 

We followed the pipeline steps with two exceptions: Due to the excellent seeing conditions for most observations (Tab.~\ref{tab:02-Observing-parameters}), the point sources in many datasets were close to the under-sampling limit $(< 0.5$\arcsec). Hence, we switched from a third- to a second-order Lanczos kernel for co-addition and resampling with SWarp \citep{swarp, Gruen14}, as it performed better near the under-sampling limit on a test sample. Secondly, the extended nebulosity of the Flame Nebula was subtracted from each tile, using a custom artifact removal algorithm (Bertin et al., in prep). The source catalog was derived from the de-nebulized image. It comprised only stars and is labeled Tile$^*$ in Fig.~\ref{fig:02-Workflow-dataset}.

\subsubsection{Source detection strategies}

We adopted a two-pronged source detection approach using both the flux-calibrated Tiles$^*$ and their associated source catalogs from \texttt{vircampype}. The pipeline uses \emph{SExtractor} \citep{1996Sextractor} for centroid-based source extraction and aperture photometry, and calibrates the stellar positions against \emph{Gaia} to ensure high accuracy. However, the central region of NGC~2024 is significantly more crowded than typical VISIONS fields, posing an edge case to the functionality of \texttt{vircampype}.

We found that \texttt{vircampype} failed to detect some sources in the dense cluster core, likely due to crowding. However, point-spread function (PSF) fitting allowed us to recover these sources, as it remains robust even in the presence of nearby or overlapping stars \citep[e.g.,][]{2017Andersen}. Using a custom \texttt{IRAF/daophot} pipeline, we applied PSF-fitting to the central region of each Tile$^*$. We generated a PSF-based source list for each dataset to maximize the detection of potential cluster members. The procedure is detailed in Appendix \ref{appendix:IRAF}.

\subsubsection{Saturated source removal}

The saturation magnitude of the VIRCAM detectors depends on the observing conditions, selected passband, and exposure time. For the source catalog provided by \texttt{vircampype}, stars brighter than a limiting magnitude determined during the processing are automatically replaced with their \emph{2MASS} measurements. We removed those stars by setting the flag \texttt{SURVEY==VISIONS}.

For stars extracted via PSF-fitting, we manually determined the saturation magnitude for each Tile$^*$ from the radial stellar profile curves of bright stars using \texttt{imexam} and used it as a filter criterion. The saturation magnitudes for each dataset, determined using both source extraction methods, are listed in Tab.~\ref{tab:03-Saturation-limits}.

\subsubsection{Concatenating the catalogs}
\label{sec:pype-psf-performance}

On average, $\sim$3,800 sources appear in both the \texttt{vircampype} catalog and the PSF-fitting source list. The former exhibit significantly higher positional accuracy when compared to the \emph{Gaia} DR3 catalog for the available subset ($n \sim 900$, depending on the dataset). Across epochs, the standard deviation of the separation between \emph{Gaia} DR3 and the VISTA source positions ranged $\sigma_{\text{$\alpha$, PSF}} \in [21.88, 114.03]$ mas and $\sigma_{\text{$\delta$, PSF}} \in [20.91, 110.72]$ mas for PSF-fitted sources, whereas \texttt{vircampype} positions were much better constrained, $\sigma_{\text{$\alpha$, vircampype}} \in [18.56, 25.85]$ mas and $\sigma_{\text{$\delta$, vircampype}} \in [16.2, 29.66]$ mas. Consequently, we adopted the \texttt{vircampype} positions for all sources present in both catalogs.

In each dataset, $\sim$300 stars were detected exclusively with \texttt{vircampype}, and $\sim$200 only via PSF-fitting (PSF-only). The former are generally faint (histogram peak at $H\approx18.7$ mag) and distributed randomly across the observed field, only sparsely populating the central region of the Flame Nebula. Their detection results from \texttt{vircampype} using a lower effective peak detection threshold than the $5\sigma$ used with \texttt{daophot}.

The PSF-only sources fall into two groups: 1) Bright sources ($H < 14$ mag), likely beyond the brightness cutoff values of \texttt{vircampype}, that we retained with the manual cuts. They are distributed across the observed field, albeit preferentially toward the cluster center. 2) Fainter sources ($H > 14$) which are either close visual double stars or located in the halos of bright stars. They were likely missed in the centroid-based source extraction due to crowding or blending. Both groups contribute substantially to the cluster member census and are therefore included in the catalog, despite their lower measurement accuracy.

\subsection{Master catalog level}
\label{sec:Method-Master}

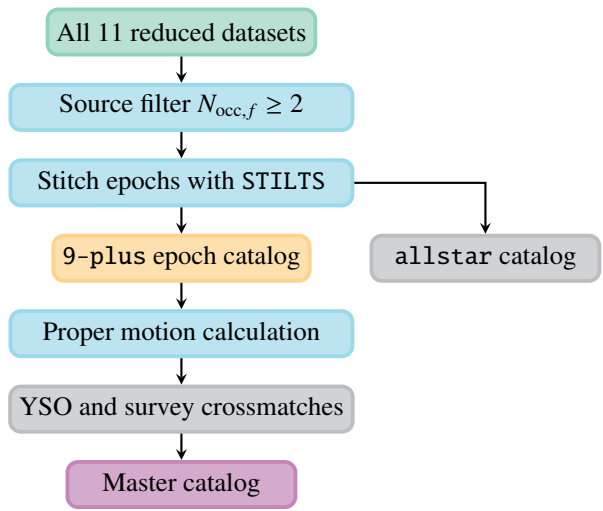
\begin{figure}[t!]

\begin{tikzpicture}[node distance=1.cm]

    \node (start) [startstop2] {All 11 reduced datasets};
    \node (nocc) [process2, below of=start] {Source filter $N_{\text{occ}, f} \geq 2$};
    \node (stitch) [process2, below of=nocc] {Stitch epochs with \texttt{STILTS}};
    \node (allstar) [product_unimportant2, below of=stitch, xshift=4cm] {\texttt{allstar} catalog};
    \node (9plus) [product2, below of=stitch] {\texttt{9-plus} epoch catalog};
    \node (ppm) [process2, below of=9plus] {Proper motion calculation};
    \node (xmatch1) [process_unimportant2, below of=ppm] {YSO and survey crossmatches};
    \node (stop) [product_important2, below of=xmatch1] {Master catalog};

    \draw [arrow] (start) -- (nocc);
    \draw [arrow] (nocc) -- (stitch);
    \draw [arrow] (stitch) -| (allstar);
    \draw [arrow] (stitch) -- (9plus);
    \draw [arrow] (9plus) -- (ppm);
    \draw [arrow] (ppm) -- (xmatch1);
    \draw [arrow] (xmatch1) -- (stop);

\end{tikzpicture}

\caption{Workflow depicting creation of the master catalog.}
\label{fig:03-Workflow-master}
\end{figure}

In the second workflow (Fig.~\ref{fig:03-Workflow-master}), the 11 processed datasets are merged into one master catalog. It comprises information on each observation, stellar positions and proper motions, auxiliary data such as YSO class, and crossmatches to other surveys.

\subsubsection{Catalog cleaning and multi-epoch source matching}

Having multi-epoch data available in all three passbands (pb) permitted an efficient removal of spurious detections from the source list using the source occurrence requirement $N_{\text{occ, pb}} \geq 2$. This approach helped eliminate spurious detections, such as peaks in the variable sky background or nebulosity that were not entirely removed during de-nebulization, since these features were not detected at the exact same location in different epochs. Only two epochs were available for the $J$ and $K_S$ filters, respectively. We used the \texttt{tmatch2} function of \texttt{STILTS} \citep{2006STILTS} with a match radius of $r = 0.5$\arcsec to match the source coordinates between these datasets. The seven $H$-band epochs were cleaned using the \texttt{tmatchn} function and iteratively treating one epoch as a reference and identifying matches in all remaining epochs.

The purpose of the master catalog was to construct a time series of position measurements, $x(t)$, for each star, which could be used for the proper motion calculation. All cleaned datasets were crossmatched using \texttt{tmatchn} ($r = 0.5$\arcsec), yielding the number of occurrences $N_{\text{occ}}$ of each source in all datasets. Two catalogs were produced: The \texttt{allstar} catalog, meaning the union of all sources with $N_{\text{occ}} \geq 2$, and the \texttt{9-plus} epoch catalog containing all sources with $N_{\text{occ}} \geq 9$. The latter was used for the proper motion calculation. The occurrence threshold was chosen as a compromise: Restricting the calculation to stars detected in all datasets (81\%) would guarantee a complete sampling of the temporal baseline for each source, but ignores the fact that many likely young stars near the cluster center were detected in the $H$ and $K_s$ filters appear to be below the detection sensitivity in the $J$-band. Moreover, even with this lower threshold, the full temporal baseline is sampled for 99.89\% of the sources.

\begin{figure*}[t]

\begin{tikzpicture}

    \node (core) [core] {\textbf{Core} 
    
    \begin{itemize}
        \item Mean positions + errors
        \item Proper motions + errors
        \item $\Delta$ (\emph{Gaia}$_{\text{ppm}}$ - IR$_{\text{ppm}}$)
        \item $N_{\text{occ}}$ + occurrence datasets
        \item Detection method (\texttt{vircampype} / PSF)
        \item Median radial velocity 
        \item Subset flags (YSO, \emph{Gaia}, ...)
    \end{itemize}};       
    
    \node (epochs) [epochs, right of= core, xshift = 5.2cm] {\textbf{Epoch data (11x)} 
    
    \begin{itemize}
        \item Observation date
        \item Positions + errors
        \item Magnitudes + errors
    \end{itemize}
    \vspace{0.2cm}
    \begin{tabular}{l|l}
    \multicolumn{1}{l}{\emph{vircampype det.}} & \multicolumn{1}{l}{\emph{PSF det.}} \\
    [2ex]
         Aperture       & Pixel positions \\
    [2ex]
         FWHM           & Pixel position errors \\
    [2ex]
         Quality flag   & Recovery fraction 
    \end{tabular}
    };     

    \node (xmatch) [crossmatch, right of= epochs, xshift=5cm] {\textbf{Crossmatch} 
    
    \begin{itemize}
        \item \emph{Gaia} DR3
        \item NEMESIS YSO catalog 
        \begin{itemize}
            \item Contamination
            \item IR class
            \item Multiplicity
        \end{itemize}
        \item SESNA (Spitzer)
        \item Meyer 1996 selection
        \item van Terwisga+2020 disk sample
    \end{itemize}}; 
    
\end{tikzpicture}

\caption{Schematic overview of the contents of the master catalog. The full table is available online at the CDS.}
\label{fig:04-Catalog-overview}
\end{figure*}

\subsubsection{Proper motion calculation}

Stellar proper motions along $\alpha$ and $\delta$ were measured using a position-error-weighted linear regression. Their uncertainties were estimated with the jackknife technique, meaning the regression model was iteratively fitted to $N-1$ of the available data points $N$, and the slope was determined for all possible permutations. The final value of the proper motion along a given position coordinate was defined as the median of all calculated slopes. For the measurement uncertainty, we used the standard deviation estimator $\hat{\sigma} = 1.4826 \cdot \mathrm{MAD}$, where $ \mathrm{MAD}$ is the median absolute deviation, assuming an underlying normal distribution \citep{ruppert2010statistics}. We used the standard correction $\mu_{\alpha*} = \mu_{\alpha} \cdot \cos(\delta_{\text{mean}})$. 

We compared the jackknife method with classical bootstrapping and found that, while the \emph{Gaia} comparisons were similarly good for both methods, the proper motion errors in both $\mu_{\alpha*}$ and $\mu_\delta$ from bootstrapping were substantially larger: Their standard deviations were roughly three times higher than the standard deviations for the errors estimated via jackknife, indicating that bootstrap errors were dispersed over a much broader range. As the jackknife procedure uses $(N-1)/N \cdot 100\%$ of the data per iteration, which in our case is $\simeq 88$--$99\%$, each bootstrap sample contains on average only $\simeq 63.2\%$ of the data. Because jackknife incorporates a larger fraction of the data in each iteration, it provides more stable and reliable error estimates, particularly for datasets with a limited number of observations. Therefore, using the jackknife method for the final catalog is justified.

For the \texttt{vircampype} catalog, where the pipeline provides correlation coefficients between the $\alpha$ and $\delta$ coordinates, we also tested the proper motion method intended for future VISIONS data releases (Meingast et al., in prep.). Including the coordinate correlations and fitting the full covariance matrix in the linear regression had no statistically significant impact on our calculated proper motion values (maximum deviation $<0.25$ mas\,yr$^{-1}$ ), their uncertainties, or the results presented in Sect.~\ref{sec:Results}. Since correlation coefficients are unavailable for the PSF-only sources and the results remain unaffected, we retained our original method, ensuring a homogeneous treatment of all sources.

\section{The master catalog of IR proper motions}
\label{sec:Results}

The workflows outlined in the previous section produce a master catalog of infrared proper motions for 6,769 sources (Fig.~\ref{fig:04-Catalog-overview}). Of these, 194 were detected and characterized via PSF-fitting, while 6,575 were taken from the \texttt{vircampype} source catalogs.

\subsection{Core and epoch columns}

The core columns of the master catalog contain the mean stellar positions, averaged over the observation epochs, the calculated proper motions, and the respective errors. For sources observed with \emph{Gaia}, they also list the difference between the \emph{Gaia} and infrared proper motions. The core columns further include information about the source detection method, the mean radial velocity ($v_{\text{LOS}}$), if measurements were available, and various subset flags, such as YSO candidate, \emph{Gaia} source, or binary candidate.

After the core columns, eleven sets of epoch columns provide information on stellar positions and magnitudes, along with their uncertainties, observation dates, and detection method-specific details. For stars detected with \texttt{vircampype}, these are the exposure time, full-width half maximum (FWHM) values, and a quality flag regarding the source extraction. For sources extracted with PSF-fitting, the columns include pixel positions, their estimated errors, and the recovery fraction of cloned sources (Appendix \ref{appendix:IRAF}).

\subsection{Crossmatches and YSO identification}

The last part of the master catalog contains details about YSO classification, radial velocities, and photometric data from the literature. For all crossmatches, we used $\alpha$ and $\delta$ positions averaged over all epochs and a matching radius of $r=2$\arcsec.

First, we matched the catalog with \emph{Gaia} DR3, yielding 1018 total matches, 899 with 5D astrometry ($\alpha$, $\delta$, $\varpi$, $\mu_{\alpha*}$, $\mu_{\delta}$).\footnote{We note that only 808 sources have $\varpi \geq 0$.} For all sources in both catalogs we calculated $\Delta (\emph{Gaia}_{\text{ppm}}-\text{IR}_{\text{ppm}})$ in the core columns. Lastly, we crossmatched with the Spitzer Legacy catalog SESNA \citep{2019SESNA} to gain information in the longer wavelength ranges.

To identify young stars, we consulted the most complete YSO catalog of the region to date \citep{2025Roquette}. In particular, we used their homogeneously calculated infrared excess index $\alpha_{\text{IR}}$ and their collated data on binarity, radial velocity, and possible contaminants. From the $\alpha_{\text{IR}}$ classification, we identified 420 YSO candidates. After removing unclassified sources and stars flagged as main-sequence, giant, or extragalactic contaminants, we defined a bona fide sample of 362 YSOs, from Class~I\footnote{\cite{2025Roquette} do not separate Class~0 from Class~I objects. Because our dataset is NIR-based and visual inspection shows no evidence for genuine Class~0 sources, we adopt their Class~0/I category as equivalent to Class~I.} to Class~III. Radial velocity information was available for 76 of these sources. When multiple measurements were present for a source, we adopted the median value and combined the associated uncertainties using inverse-variance weighting.

We investigated the 108 sources detected via PSF-fitting that were not in the YSO crossmatch. Crossmatching those sources with Spitzer and allWISE yielded 29 matches, and only two of those might be YSOs $(4.5-8)~\mu$m $> 0.5$ \citep{2014Koenig,2015Spezzi}. All other stars, only eleven of which are located in the cluster region, are not detected by those two surveys. We conclude that our YSO sample is representative and includes most young cluster members. 

We also crossmatched to the photometric survey of NGC~2024 by \cite{1996Meyer} and a disk population study based on this survey \citep{Terwisga2020}. We discuss the YSO fraction of their surveys and the disk population of the cluster in Appendix \ref{appendix:Disks}.

\subsection{Comparison to Gaia}

\begin{figure}[ht]
    \centering
    \includegraphics[width=\linewidth]{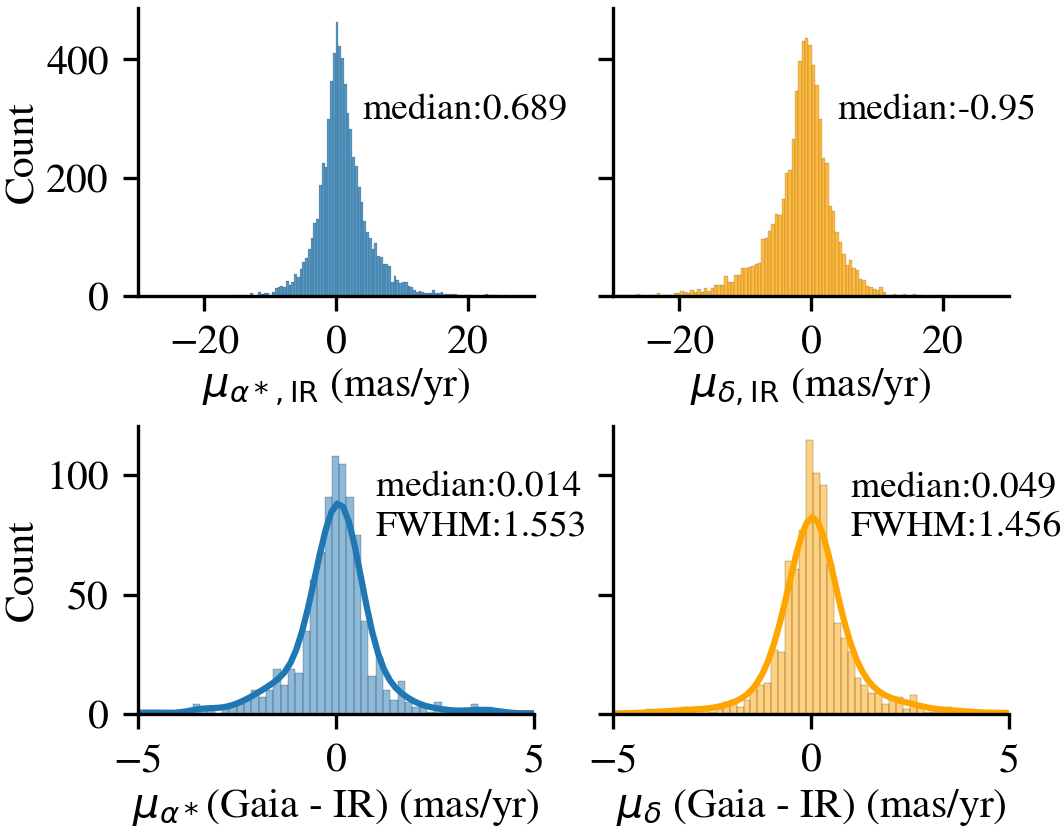}
    \caption{Infrared proper motion histograms for all 6,769 sources (top) and deviation from \emph{Gaia} proper motions for 899 shared sources (bottom).}
    \label{fig:05-ppm-hist}
\end{figure}

\begin{figure}[h!]
    \centering
    \includegraphics[width=0.99\linewidth]{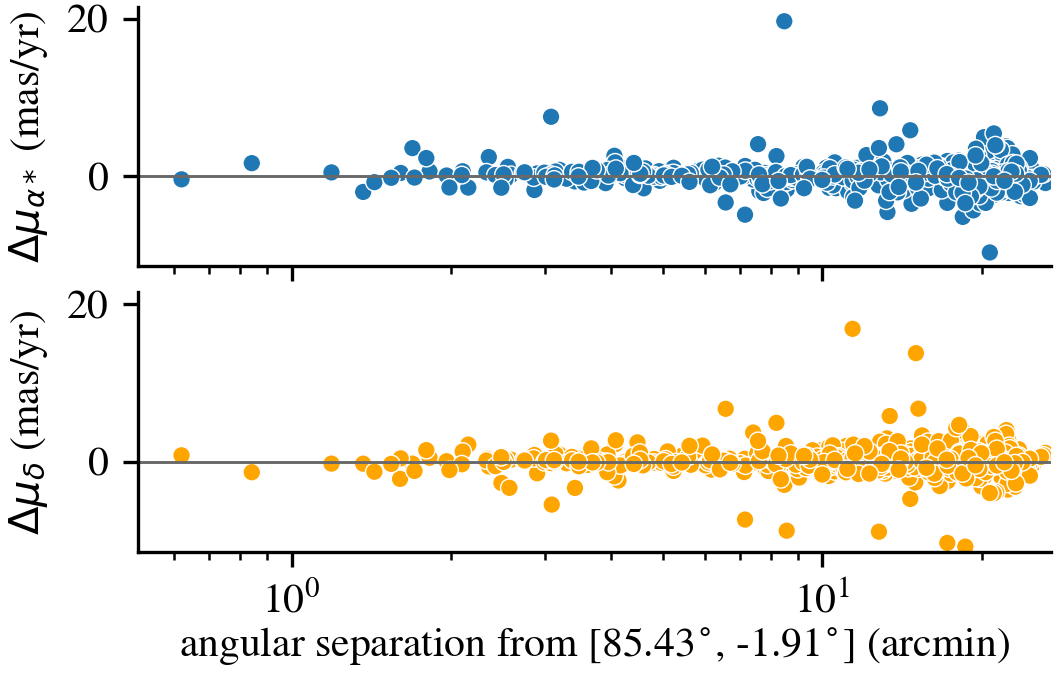}
    \caption{Difference between \emph{Gaia} and infrared proper motions versus angular distance from the projected cluster center. The log-scale of the x-axis highlights the lack of \emph{Gaia} sources near the center.}
    \label{fig:07-ppm-center-diff}
\end{figure}

\begin{figure}[h!]
    \centering
    \includegraphics[width=\linewidth]{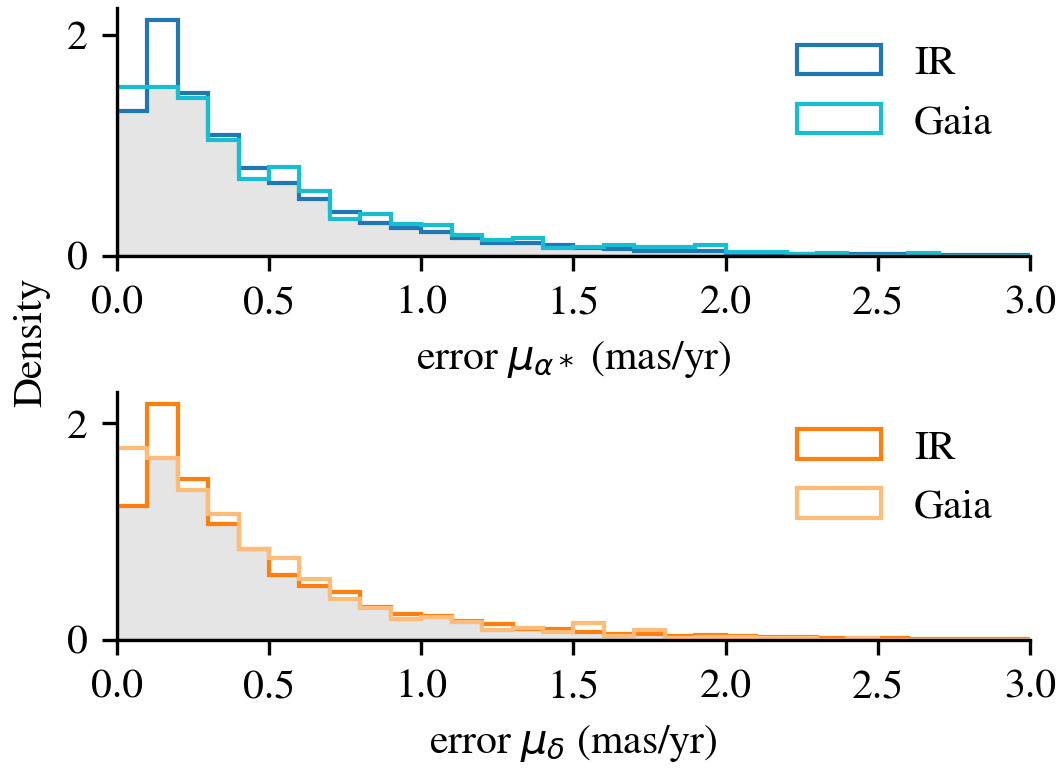}
    \caption{Comparison between \emph{Gaia} and IR proper motion errors. The gray shaded area indicates the overlap between the two distributions.}
    \label{fig:06-Gaia-comp-eppm}
\end{figure}

The proper motion histograms for all sources in the master catalog are shown in the top row of Fig.~\ref{fig:05-ppm-hist}.
The catalog includes sources with proper motions up to $\pm 20$ mas\,yr$^{-1}$, which likely are foreground stars. Color-color and color-magnitude diagrams show that the bulk of stars in the catalog appear to be galactic background objects, with extinctions up to $A_V\sim10$ mag. 

The bottom row of Fig.~\ref{fig:05-ppm-hist} shows the difference between the proper motions measured by \emph{Gaia} DR3 and those determined from infrared data. The difference distributions are centered almost at zero, with very narrow FWHM values of around 1.5 mas\,yr$^{-1}$. This indicates an excellent agreement between the \emph{Gaia} and infrared measurements, with no prominent systematic offsets present for the catalog on the whole.

\begin{figure*}[!htbp]
    \centering
    \includegraphics[]{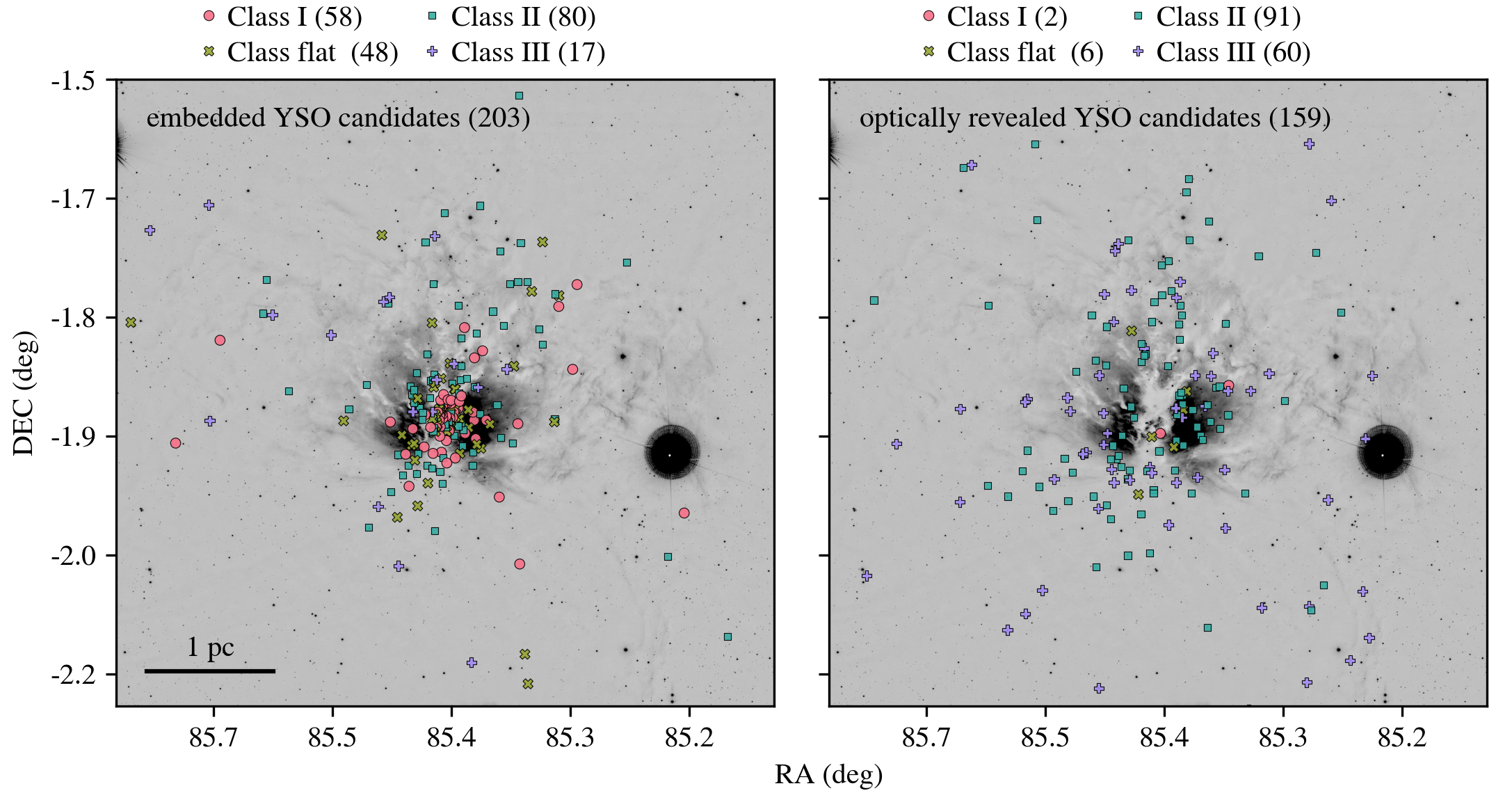}
    \caption{Positions of the 362 identified bona fide YSO candidates atop the inverted VHS $J$-band cutout. The stars are divided into still embedded sources detected in the NIR, but not by \emph{Gaia}, and optically revealed YSOs. The scale bar shows the angular scale of 1 pc at a distance of 420 pc.}
    \label{fig:07-YSO_classes_Gaia-division}
\end{figure*}

We note that for a shared subset between \emph{Gaia} and the 98 sources determined only with PSF-fitting, we see a broader scatter of $\mathrm{FWHM}(\mu_{\alpha*})\approx 2.327$ mas\,yr$^{-1}$ and $\mathrm{FWHM}(\mu_{\delta})\approx 2.483$ mas\,yr$^{-1}$. The broadening is likely a direct result of the poorer positional accuracy of the PSF-fitting versus the performance of \texttt{vircampype} source extraction, which we observed during the direct comparison with the available \emph{Gaia} data. Additionally, the median of the difference distributions in both variables is offset by about $0.2$ mas\,yr$^{-1}$. We suspect this shift originates from one or two observation epochs, where the PSF-fitting yielded the worst positional accuracy relative to \emph{Gaia} (see upper limits in Sect.~\ref{sec:pype-psf-performance}). We stress that this shift equals a velocity of 0.38 km\,s$^{-1}$ at $d\sim400$ pc, which is significantly smaller than the differences discussed later.

We examined whether the deviation between \emph{Gaia} proper motions and those derived in this work varies systematically with distance from the 2D-projected cluster center in Fig.~\ref{fig:07-ppm-center-diff}. Such a dependence could indicate, for instance, less accurate \emph{Gaia} measurements in the cluster center due to the high optical extinction. We find no linear trend between the deviations in proper motion and radial distance. This also holds when applying different binning schemes (equal radii, equal area, or equal source number). The log-scale in Fig.~\ref{fig:07-ppm-center-diff} highlights the lack of \emph{Gaia} measurements near the cluster center and the increasing source number towards the outskirts of the nebula.

In Fig.~\ref{fig:06-Gaia-comp-eppm}, we compare the measurement errors of \emph{Gaia} DR3 with our results. It is evident from the significant overlap of the two distributions that the errors are of the same order of magnitude. We conclude that our proper motions are of a similar quality to the ones produced by \emph{Gaia} for this region.

\subsection{Embedded vs. optically revealed YSOs}
\label{sec-YSO analysis}

We show the positions of the YSO candidates atop an inverted VHS \mbox{$J$-band} cutout in Fig.~\ref{fig:07-YSO_classes_Gaia-division}. Of our 362 bona fide YSO sample, 203 appear only in our infrared catalog, which strongly suggests they are embedded sources. The other 159 stars are already optically revealed as they were detected with \emph{Gaia}. This means that with our catalog, we gain proper motions for more than half ($\sim 56\%$) of the YSO candidates identified in the direction of NGC~2024. Moreover, the embedded population includes 58 out of 60 Class~I objects, along with 89\% of the Class~flat YSOs, marking a $>13$-fold increase in kinematic information on the youngest two YSO classes. Class~II objects are almost evenly distributed between the embedded and optically revealed groups, and about 78\% of Class~III stars are optically revealed. 

The covariance matrices of the groups show that the embedded group is more centrally concentrated, with $1\sigma$ positional spreads of $\sim4.8$\arcmin $\times~5.0$\arcmin, compared to a larger and more elongated $\sim5.9$\arcmin $\times~7.1$\arcmin for the group detected by \emph{Gaia}.

\section{Discussion}
\label{sec:Discussion}

\begin{figure*}[ht!]
    \centering
    \includegraphics{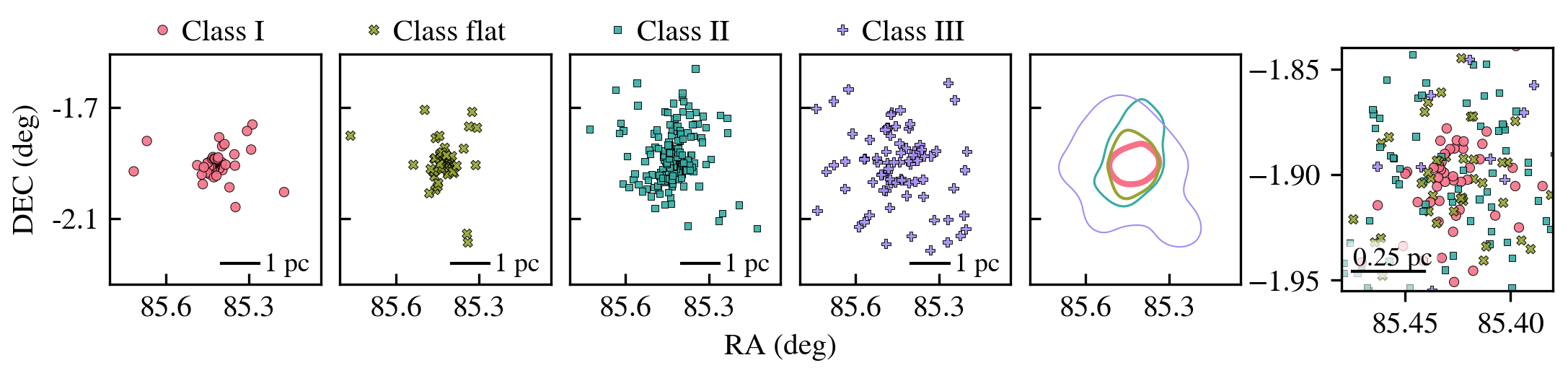}
    \caption{Panels $1-4$ show the positions of the identified YSO candidates, separated by evolutionary class. Panel 5 shows kernel density estimates (KDEs) at the 25\% contour level, and panel 6 provides a zoom-in of the cluster center. The contour linewidths decrease with increasing evolutionary stage, reflecting the YSO classes ordered by age. The scale bars are given for a distance $d\sim 420$ pc.}
    \label{fig:09-YSO_classes_Gaia-kde}
\end{figure*}

The positions and stellar kinematics of YSOs encode the combined influence of both their coupling to their gaseous environment and their cluster assembly history. According to recent simulations \citep[e.g.,][]{2025Laverde}, the way in which a cluster forms -- whether through the hierarchical subcluster mergers or more monolithic growth -- imprints distinct spatial and dynamical signatures on its stellar population. Similarly, the timescale over which stars decouple from their natal gas may be reflected in observed differences in the kinematic profiles of different YSO classes \citep[e.g.][]{2009Proskow, 2018Sills}. It is worth comparing the stellar velocities to those of the surrounding gas to pin down the gas-star coupling duration. In the following, we analyze NGC~2024 concerning these questions, using the newly gained and collected information on its stellar kinematics. For this discussion, we assume that the identified YSO candidates \citep{2025Roquette} are members of NGC~2024. We support this assumption using Fig.~\ref{fig:07-YSO_classes_Gaia-division}, and the calculated covariance matrices, which show that the YSO candidates -- particularly the youngest -- are concentrated toward the 2D-projected cluster center. While some contamination cannot be ruled out, most sources are likely genuine cluster members.

\subsection{Probing the assembly history of NGC~2024}

Previous studies \citep{2014Getman, 2018Getman} identified an age gradient within the central $\sim1$ pc of NGC~2024, with younger sources near the cluster center and older sources toward the outskirts, based on 121 identified members. The discovery paper also lists eight scenarios for the appearance of such intra-cluster age gradients. They can be broadly grouped into: 
\begin{enumerate}
    \item More recent star formation in the cluster center (more dense material; accelerated star formation rate; late, central formation of massive stars and companions)
    \item Outward movement of older stars (radial drift; dynamical heating; expansion of older subclusters)
    \item Inward movement of young stars (late infall of filaments with already formed stars; subcluster mergers)
\end{enumerate}

No scenario was favored in their study, likely in part because kinematic data for the cluster members were not previously available to evaluate the dynamical aspects of these formation pathways. As noted by \cite{2003Bate}, the typical stellar velocity dispersion in young clusters ($\sim1-5$ km\,s$^{-1}$), found in simulations and observations, allows stars to traverse distances comparable to the initial cloud radius ($\sim 0.1-0.2$ pc) over typical star formation timescales of $\sim10^5$ yr. Thus, positions are insufficient to evaluate cluster assembly histories.

Some formation scenarios of \cite{2014Getman} involve subclusters, whereas others fit a more monolithic assembly. The question of monolithic versus hierarchical cluster formation is also central in simulations. They agree on an early hierarchical assembly, but diverge on how long remnants of such substructures remain observable. For instance, \cite{2025Laverde} report a low-level kinematic substructure reminiscent of subcluster merger events still present after $2.5$ freefall times of a cluster, even though spatial substructures have already disappeared by then. For cloud masses on the order of $10^4 M_{\odot}$, this would correspond to $\sim$2.5 Myr. 

In contrast, simulations by \cite{2018Sills}, motivated by observed initial conditions \citep{2014Getman_many}, showed a collapse of a cluster from an initially substructured state into a spherical, centrally dominated cluster already within $0.5-1$ Myr. This collapse could also cause a variation in the velocity dispersions as a function of YSO age \citep[Fig.~12,][]{2018Sills}.

With our homogeneous proper motion information across most YSO evolutionary classes that appear in the cluster, we revisit and discuss possible formation scenarios of NGC~2024. We do not focus on the first group of scenarios listed in \cite{2014Getman}, as they cannot be directly addressed using the kinematic information for the young stars we inferred in this study.

\subsubsection{Position-age gradient and internal motion}
\label{sec:age-grad}

We first investigate whether the previously reported age gradient from younger, central stars to older stars at larger radii is present in our dataset. Since we do not measure absolute stellar ages, we use YSO classes as proxies. Of the 121 sources reported in \cite{2014Getman}, 106 are included in our catalog; the 15 non-recovered sources have \emph{2MASS} $H$-band magnitudes brighter than 12~mag, exceeding our saturation limit. Additionally, three sources are classified as non-members, and nine are flagged as main-sequence contaminants in \cite{2025Roquette}.

In Fig.~\ref{fig:09-YSO_classes_Gaia-kde}, we analyze the position of the cluster members divided by YSO class. Their corresponding covariance and spatial dispersion parameters are listed in Tab.~\ref{tab:YSO-covariances}. Most Class~I and flat objects are located near the central nebulosity. Class~II objects occupy a larger range along the declination axis, but are still packed densely. Class~III objects generally display a more scattered behavior. The covariance analysis of the YSO classes shows a clear trend from younger, more tightly clustered YSOs to older, more spatially dispersed YSOs.

The KDE is elongated in the N-S direction for the Class~II and flat sources, also when considering different density thresholds. Likewise, the 1$\sigma$ ellipses of Classes flat and II are elongated in the N–S direction ($\sigma_{\delta}$/$\sigma_{\alpha} > 1.2$), while those of Class 1 stars are elongated E–W instead ($\sigma_{\delta}$/$\sigma_{\alpha} < 0.7$). The dispersion of Class~III is nearly isotropic with only a slight N–S elongation. The dominant N–S elongation of the Class~II and flat-spectrum population (62\% of YSOs) agrees with previous findings \citep{2008Meyer} and suggests that stellar positions reflect the shape of the dense central nebulosity of the Flame Nebula, which itself follows the large-scale structure of the parent cloud \citep{2024Hacar}. This alignment may reflect the coupling of these YSOs to the gas, while the more evolved Class~III objects are already dispersed. If the “wick” of the Flame Nebula is approximately cylindrical, the distribution of Class~II and flat-spectrum sources may also trace a cylindrical three-dimensional cluster structure. The absence of the N-S elongation trend in Class~I objects could be due to their young age, extinction limiting our sample, or them experiencing different dynamical or environmental conditions than the older classes.

The spatial distribution of the different YSO classes suggests a mild age segregation between younger sources in the cluster center and older sources that populate the outskirts and parts of the center, similar to the age gradient reported by \cite{2014Getman}. While they found this gradient within the central parsec, our $\sim$3 times bigger sample reveals differences in YSO distributions beginning within 1~pc and extending to at least 2~pc from the cluster center. However, our sample appears somewhat mixed in the cluster center, with different populations overlapping. Only in the central $\sim0.25$ pc, seen in the right panel of Fig.~\ref{fig:09-YSO_classes_Gaia-kde}, do we definitely see Class~I objects occupying a tight space surrounded by older Class~flats and a few older sources.

We transformed the proper motions into the Local Standard of Rest (LSR) and corrected them for the cluster’s bulk motion to connect the YSO positions and motions. The resulting internal kinematics of the YSO classes are displayed as vector diagrams in Fig.~\ref{fig:09-YSO_vector_plot_LSR}. They show that Class~I and flat sources seem to expand isotropically. Higher proper motion Class~II stars show possible preferential motion along the N–S axis of the Flame Nebula’s dark lane, suggesting an influence from the gas reservoir not seen in younger populations. Class~III sources show no preferred expansion trend. We observe no indications of subclusters or their remnants in our YSO sample.

We tested whether the YSOs follow a Larson-type velocity–size relation \citep{1981Larson,1987Solomon}, but we do not find clear evidence for such a behavior in our sample. The finding is consistent with indications that the Larson relations may not hold for small spatial scales, where molecular cloud column densities can no longer be described as constant \citep{2010Lombardi} (see Appendix~\ref{appendix-Larson} for details).

\subsubsection{Kinematic properties across YSO classes}
\label{sec:vel}

\begin{figure}[t]
    \centering
    \includegraphics[width=\linewidth]{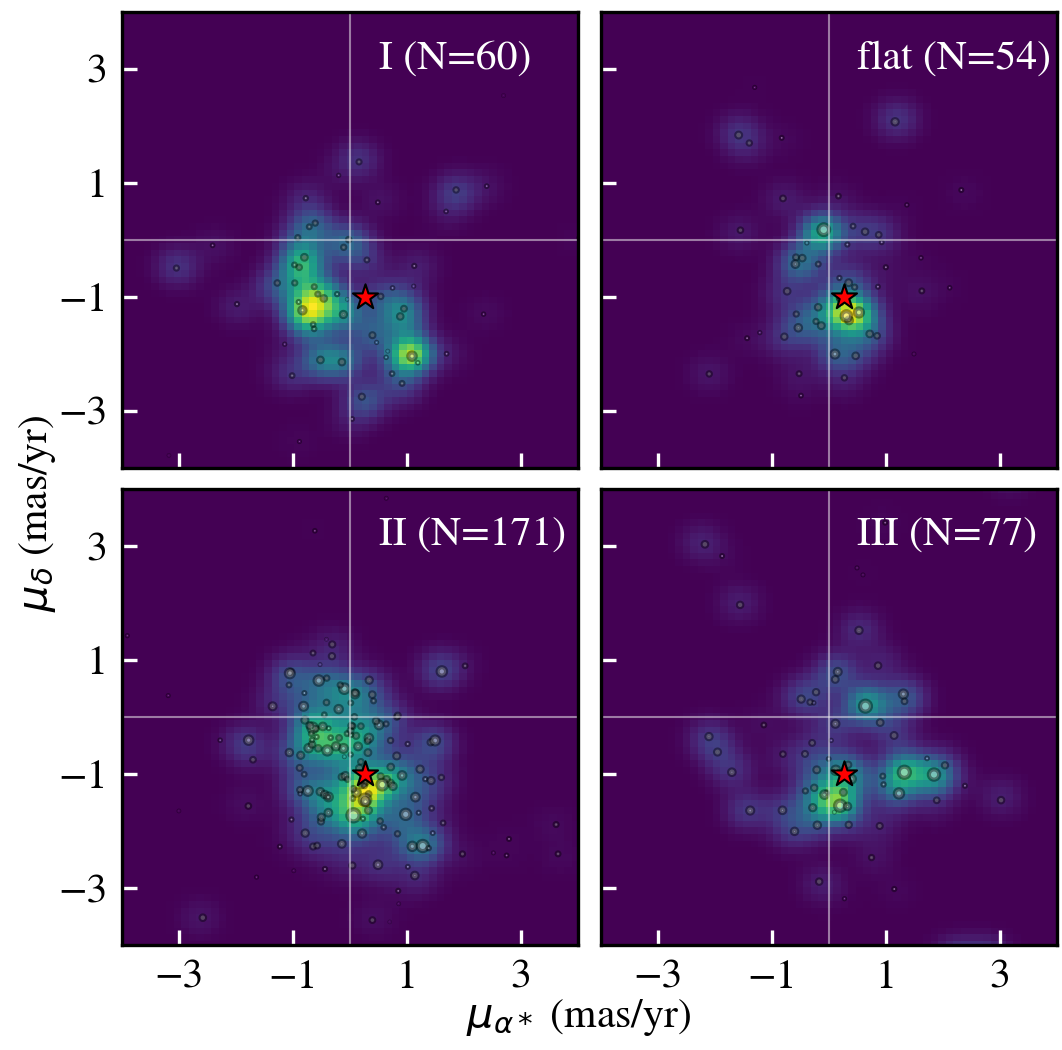}
    \caption{2D proper motion histograms of the YSO classes, created with 500 bins and smoothed with a Gaussian with $\sigma=2$ bins/pixel. The black circles show the actual proper motions, and the marker size corresponds to the weight $w$ of a data point. The red star marks the mean YSO proper motion $(0.26, -1.01)$ mas\,yr$^{-1}$ .}
    \label{fig:12-YSO-heatmaps}
\end{figure}

With the added proper motions of the optically embedded objects, we can create a more complete picture of the velocity space that the cluster occupies than previous, \emph{Gaia}-based studies. Figure~\ref{fig:12-YSO-heatmaps} presents 2D velocity histograms for the different YSO classes, where each source is weighted by its proper motion errors $w = (e_{\mu_{\alpha*}}^2 + e_{\mu_{\delta}}^2)^{-1}$. The weights are summed within each bin, and the binning and smoothing parameters were determined empirically to balance the resolution of velocity structures against noise from sparse sampling.

As can be seen, there is generally a large scatter of the YSO proper motions across the parameter space, regardless of evolutionary class. This finding is in agreement with previous studies. For example, \cite{2024Zerjal} identified 62 cluster members, which are difficult to discern as an overdensity in proper motion space while appearing much more concentrated in positional space. Despite containing hundreds of members, the cluster is absent from most large-scale cluster catalogs \citep{2023Perren}, likely due to the large velocity dispersion of its members and the fact that its embedded core is not detected by \emph{Gaia}.

Still, the highest density peaks in the distributions of the Class~flat, II, and III populations are near the mean cluster velocity marker at $(0.26 -1.01)$ mas\,yr$^{-1}$. The same is not true for the Class~I objects, where the peak is around $(-0.67,-1.16)$ mas\,yr$^{-1}$. The one-dimensional shift between the Class~I peak and the mean cluster proper motion corresponds to $\Delta \mu_{\alpha*} = 0.93$ mas\,yr$^{-1}$, or around 1.85 km\,s$^{-1}$. We note that the reported systematic shift of ca. 0.2 mas\,yr$^{-1}$ we found between \emph{Gaia} and PSF-only proper motions cannot produce this difference. It also appears when only using sources in the \texttt{vircampype} source catalog. However, we could not find any spatial correlation between the Class~I stars producing the velocity overdensity. We also investigated the secondary peaks in the velocity distribution of Class~flat and Class~II, toward $(-0.35,-0.35)$ mas\,yr$^{-1}$, but found no spatial correlations. 

To test whether the Class~I proper motion is statistically different from that of the rest of the cluster, we applied mean-based permutation tests, 2D Kolmogorov–Smirnov (KS), and energy-distance tests to the ($\mu_{\alpha*}$, $\mu_{\delta}$) distributions of different YSO classes. The permutation test did not detect significant differences in mean velocities ($p$-values $>0.17$). The 2D KS-test $p$-values were all larger than the corrected threshold of $0.05/N_{\rm tests} = 0.009$. The Class~I vs Class~III 2D KS test had the lowest value ($p= 0.04$). The energy-distance test, which is sensitive to differences in distribution shape as well as mean, identified only one pair (Class~III vs flat) as marginally distinct $(p = 0.033)$, with Class~III vs Class~I close to the 0.05 threshold $(p = 0.062)$. All other pairs are consistent with sharing the same distribution. Overall, we find that the kinematic distributions of YSO classes in NGC~2024 are statistically indistinguishable, even though the distribution of Fig.~\ref{fig:12-YSO-heatmaps} seems to suggest substructures at first glance. The marginal difference between Class~III and I/flat sources found in different statistical tests suggests that the oldest population is more dynamically dispersed than the embedded stars, but this is not a dominant effect. We conclude that all classes share a common velocity distribution within $\leq 2$ km\,s$^{-1}$, indicating rapid dynamical mixing or formation in a relatively coherent structure. The lack of spatial correlation for all investigated density peaks also supports this scenario.

\begin{figure*}[t!]
    \centering
    \includegraphics{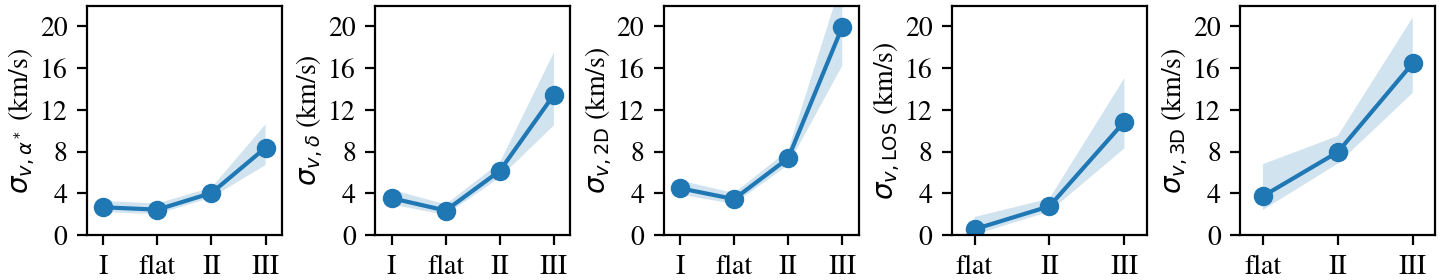}
    \caption{
     Evolution of the velocity dispersions of the different YSO classes. The blue markers show the posterior median velocity dispersion alongside the 95\% credible intervals. Note that insufficient Class~I data are available for the two rightmost panels.}
    \label{fig:10-velocity-params}
\end{figure*}

To gain further insight into the cluster dynamics, we investigated the 1D, 2D, and 3D velocity dispersions as a function of YSO class. To determine the intrinsic velocity dispersion, we modeled the underlying velocity distribution of each YSO class. Given the limited number of observations per class, we adopted a Bayesian framework that allows for a principled treatment of uncertainty. Details about the full methodology are given in Appendix~\ref{appendix:Bayesian-model}. In brief, the velocity distribution of young stellar populations is typically well described by a multivariate normal distribution $\mathcal{N}$ with the covariance matrix $\mathbf{S}$ representing the intrinsic dispersion. Accounting for heteroscedastic Gaussian measurement uncertainties $\mathbf{C}_i$ of each star $i$, the observed velocities can be modeled as the convolution of the intrinsic multivariate normal distribution with the observational error distribution, resulting in a likelihood of the form
\begin{align*}
\mathbf{v}_i \sim \mathcal{N}\!\left(\boldsymbol{\mu},\, \mathbf{S} + \mathbf{C}_i\right).
\end{align*}
We used uninformative priors so that the data dominate the posterior and the inferred dispersions reflect the observed kinematics. We sampled the posterior distribution of the intrinsic 1D$-$3D velocity dispersions for each group using an MCMC sampler \citep{10.5555/2627435.2638586}. Figure~\ref{fig:10-velocity-params} shows the resulting median velocity dispersions along with their 95\% credible intervals, based on $10,000$ posterior samples. Tangential velocities were calculated as $v_t \text{( km\,s$^{-1}$)} = 4.74057 \cdot \mu~\text{(mas\,yr$^{-1}$)} \cdot d\text{~(kpc)}$ assuming $d=420$ pc for stars without \emph{Gaia} distances. Radial velocities are available for 76 of the 362 YSOs, resulting in smaller samples for the calculations of the two rightmost panels in the figure. Since only two radial velocity measurements are available for Class~I objects, this class is excluded from the analysis of $\sigma_{v,\text{LOS}}$ and $\sigma_{v,\text{3D}}$.

The velocity dispersions for Class~I and Class~flat range between $2.4-4.5$ km\, s$^{-1}$, overlapping but slightly higher than the $1-3$ km\,s$^{-1}$ range reported for $1-5$ Myr old star-forming regions \citep{2019Kuhn} and from simulations \citep[e.g.,][]{2003Bate}. The only deviation is the low dispersion of $\sigma_{v,\text{LOS}}(\text{Class flat}) =0.57^{+1.18}_{-0.53}$km\, s$^{-1}$. Only seven RV sources are available for this measurement, resulting in a broad credible interval that reflects its high uncertainty. The velocity dispersions of Class~II are slightly higher, between $4-7.4$ km\, s$^{-1}$, while the most evolved YSOs, Class~III, show considerably higher dispersions, indicating either more dynamical evolution or possible outliers in this population.

Across all velocity dimensions, we observe the general trend of increasing velocity dispersion with evolutionary class, from Class~flat to Class~III. However, Class~I sources depart from this trend, with velocity dispersions either equal to or exceeding those of Class~flat. In 1D, Class~I has a significantly higher velocity dispersion $\sigma_{v,\delta}(\text{Class I}) = 3.55^{+0.82}_{-0.61}$km\, s$^{-1}$ compared to $\sigma_{v,\delta}(\text{Class flat}) =2.36^{+0.58}_{-0.41}$km\, s$^{-1}$, with their credible intervals directly abutting. The same trend is observed in 2D, although less pronounced. The dispersion ranges are $\sigma_{v,\text{2D}}(\text{Class I}) = 4.50^{+0.75}_{-0.61}$km\, s$^{-1}$ and $\sigma_{v,\text{2D}}(\text{Class flat}) =3.46^{+0.57}_{-0.46}$km\, s$^{-1}$, with a slight overlap of the credible intervals. 

To assess the influence of potential outliers, we repeated the inference using a multivariate Student-t likelihood function with degrees of freedom $\nu \in [1, 30]$ (see Fig.~\ref{fig:14-velocity-params-tdist}). Smaller $\nu$ values correspond to heavier tails and greater robustness against outliers, while $\nu \rightarrow \infty$ recovers the Gaussian likelihood. We find that the trends of the velocity dispersions shown in Fig.~\ref{fig:10-velocity-params} are preserved for all $\nu > d$ of the respective velocity variable. Further details are given in Appendix \ref{appendix:Bayesian-model}.

This result is notable because the otherwise approximately linear increase in velocity dispersion with YSO evolutionary stage from Class~flat to Class~III is interrupted by the higher dispersions of Class~I sources in all velocity dimensions where measurements are available. Regarding higher velocity dispersions of younger YSO classes, \cite{Sullivan_2019} investigated YSO velocities in the L1688 cluster in Ophiuchus. They saw a trend for higher 1D velocity dispersion in proper motion and radial velocity for Class~I/flat YSOs in the cloud core, compared to Class~II/III dominated samples in the lower extinction periphery. In radial velocity, the difference between 32 embedded Class~I or flat-spectrum sources was higher than that of optical surveys at the 2$\sigma$ level. In our data, we only observe qualitative differences between Class~I and Class~flat, but we do not find the same trend they reported when comparing Class~flat with Class~III sources in the 2D, $v_{\text{LOS}}$, and 3D velocity distributions. More radial velocity measurements for Class~I objects are needed to thoroughly compare the findings in NGC~2024 to those in L1688.

\subsubsection{Comparison of different formation theories}
\label{sec:formation-comp}

Connecting our analysis to the proposed star formation scenarios for NGC~2024, the findings of Sect.~\ref{sec:age-grad} agree with an outward drift of older stars, which can be directly seen from the stellar positions and directions of the proper motion vectors (Figs.~\ref{fig:09-YSO_classes_Gaia-kde} and \ref{fig:09-YSO_vector_plot_LSR}). Considering the higher 3D velocity dispersions reported for Class~II and III objects, the outward movement of the older stars could be at least in part driven by N-body interactions, in accordance with simulations \citep{2003Bate}.

There is no evidence from our analysis that young stars are currently moving toward the core of NGC~2024 due to the infall of filaments with already formed stars. Furthermore, we find no statistically significant spatio-kinematic subclusters in NGC~2024. This result contradicts the suggestion of prolonged ($\sim2$ Myr) subcluster expansion or hierarchical mergers causing the age gradient \citep{2014Getman}. It also disagrees with the simulations by \citep{2025Laverde} as no statistically significant kinematic imprint of subcluster mergers seems to appear in our analysis despite the young cluster age (Sect.~\ref{sec:vel}).

Instead, our results align with the findings of \cite{2019Kuhn}, who showed that evidence of hierarchical formation is no longer detectable in star-forming regions aged $1-5$ Myr. Our analysis even strengthens their conclusion, as it also incorporates the kinematics of the embedded cluster population, whereas their study was limited to \emph{Gaia} proper motions of more evolved objects.
Our findings would therefore support either a relatively coherent, monolithic formation of NGC~2024, or a rapid and complete dynamical mixing of any initial substructure on timescales $< 1$ Myr, as proposed by \cite{2018Sills} and also detailed in \cite{2019Kuhn}. Their simulation matches our science case for various reasons: First, the filament structure of the Flame Nebula would be well-represented, as they modeled their initial structure as a chain of subclusters. However, the simulation is considerably more massive (3000 $M_\odot$) than our cluster. Secondly, their simulations can produce rising velocity dispersions for younger YSOs, which also depend on the gas mass of the cluster. While our 1D and 2D velocity profiles as a function of relative YSO age differ from theirs in terms of absolute numbers, the qualitative trend of changing velocity dispersion across YSO classes is visible in our sample \citep[e.g.][Fig.~19]{2019Kuhn}. The rapid collapse and rebound of the cluster could account for a higher velocity dispersion of the Class~I population, while remaining consistent with the finding that all proper motions originate from the same distribution. We note, however, that our Class~III objects deviate from the flattened velocity dispersion trend seen in their simulations. Lastly, \cite{2018Sills} state that intra-cluster age gradients \citep{2014Getman} could be preserved for up to 2 Myr in their simulations, accompanied by gradual inward mixing of older stars. This agrees with the results presented in Fig.~\ref{fig:09-YSO_classes_Gaia-kde} and the estimated age of NGC~2024 of $<2$ Myr.

We note that the kinematic structure we observe in Fig.~\ref{fig:10-velocity-params} could also be a projection effect arising from the viewing angle and the elongated shape of the cluster, as suggested by \cite{2009Proskow}. If it is real, however, it could indicate a supervirial state of less evolved objects in the cloud core reported by \cite{Sullivan_2019}. In our data, only the youngest sources (Class~I) show an elevated velocity dispersion, while the Class~flats produce a local minimum. This could mean that the cluster had just rebounded following its initial collapse when the youngest YSOs in our sample were formed \citep{2019Kuhn, 2018Sills}.

\subsection{Connection to gas velocity}

\begin{figure*}[t!]
    \centering
    \includegraphics{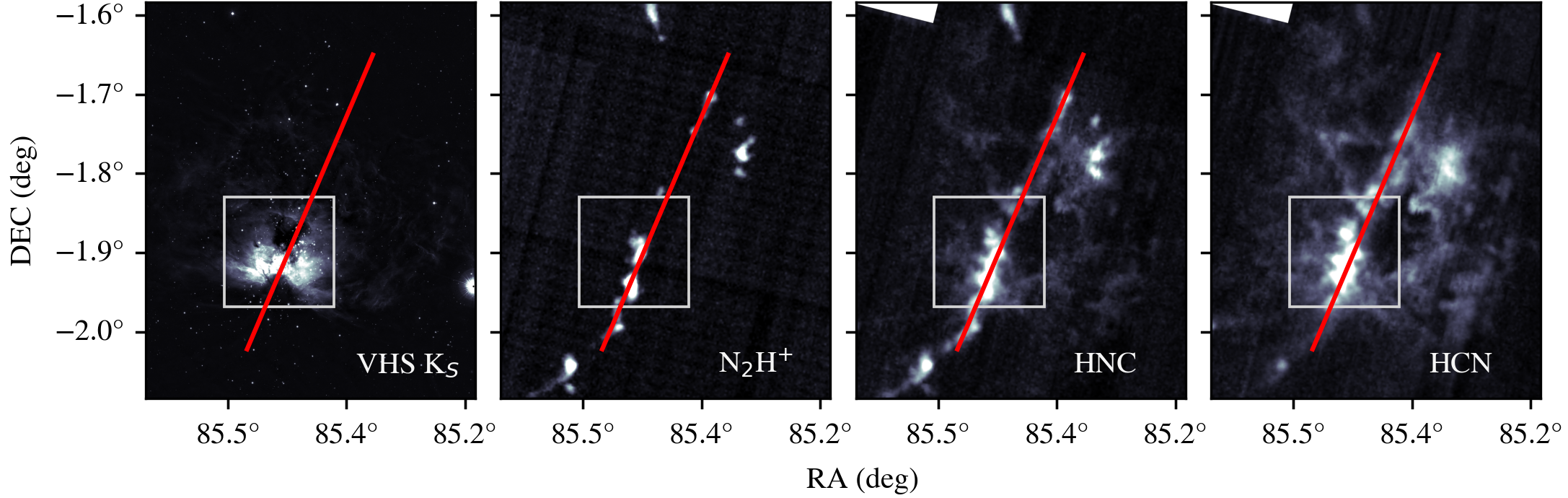}
    \caption{Location of the center of NGC~2024 (white square) and the PV-diagram path (red line) for Figs.~\ref{fig:13-gas-RV-CO} and \ref{fig:14-pv-diagrams}. The panels show a cutout of the VHS $K_S$-band observation, and the maximum gas temperature for the $(1-0)$ molecular transitions of N$_2$H$^+$, HNC, and HCN, respectively.}
    \label{fig:12-gas-RV-path}
\end{figure*}

To assess the star-gas coupling of NGC~2024, we compare the radial velocity subset of 76 bona fide YSOs with gas velocities of the Flame Nebula region using position-velocity (PV) diagrams along the path shown in Fig.~\ref{fig:12-gas-RV-path}. We use the gas tracers N$_2$H$^+$, HNC, and HCN, measured in the IRAM 30m ORION-B Large Program \citep[][PIs: M. Gerin and J. Pety]{2017Pety}, and $^{12}$CO$(3-2)$ measurements of the ALCOHOLS survey \citep{2022Stanke}. he ORION-B data have an angular resolution of 31\arcsec and a velocity resolution of 0.5 km\,s$^{-1}$, while the ALCOHOLS data provide 19\arcsec and 0.25 km\,s$^{-1}$, respectively.

\begin{figure}[t!]
    \centering
    \includegraphics[width=\linewidth]{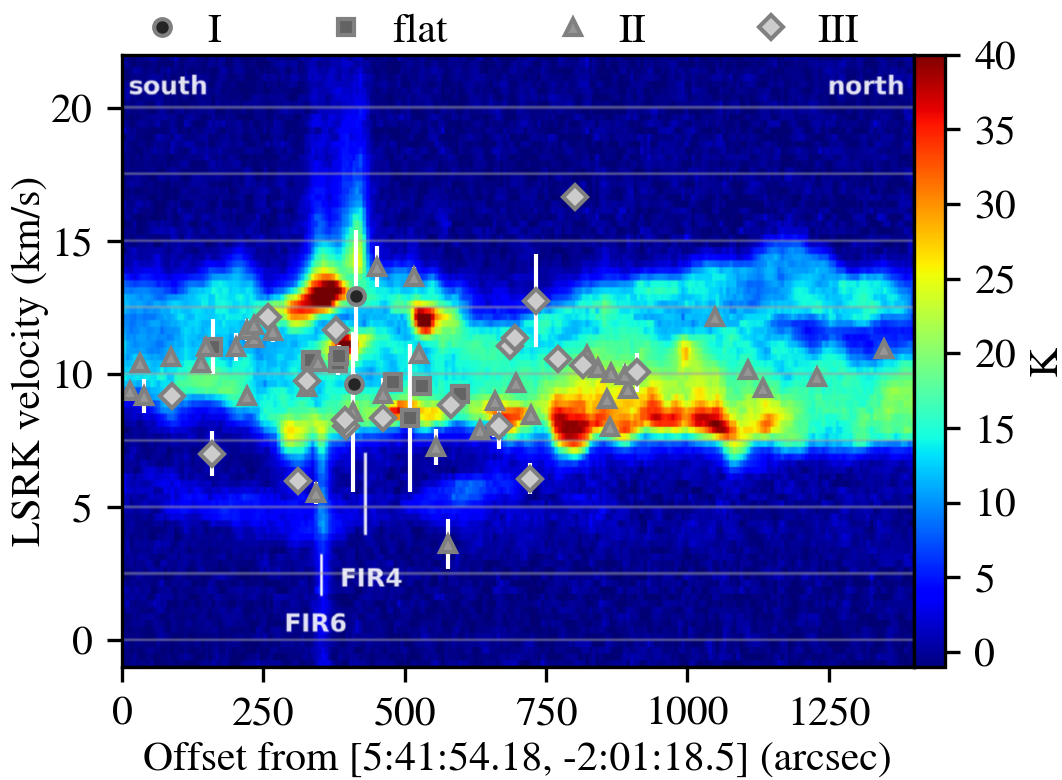}
    \caption{PV-diagram of $^{12}$CO$(3-2)$ from the ALCOHOLS survey, with YSOs from our radial velocity sample overlaid. The gas diagram is taken from \cite{2022Stanke}, Fig.~25, with the authors' permission.}
    \label{fig:13-gas-RV-CO}
\end{figure}

\begin{figure}[t!]
    \centering
    \includegraphics[width=\linewidth]{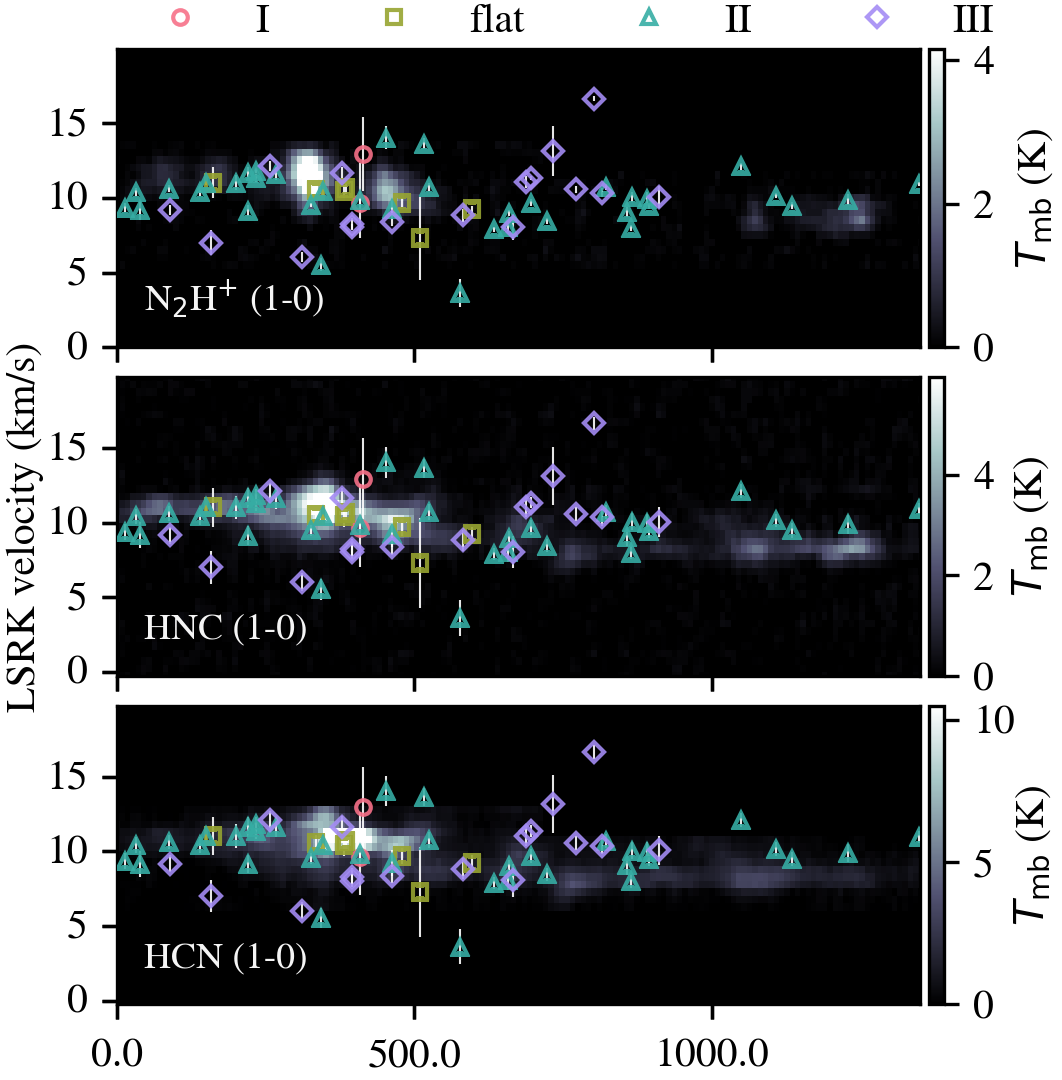}
    \caption{PV-diagrams of the $(1-0)$ molecular transitions of N$_2$H$^+$, HNC and HCN \citep{2017Pety}, overlaid with the stars in our radial velocity sample.}
    \label{fig:14-pv-diagrams}
\end{figure}

We observe a good agreement between Class~I, Class~flat, and most Class~II and Class~III sources and the $^{12}$CO gas in Fig.~\ref{fig:13-gas-RV-CO}. Class~III objects seem more scattered than their younger counterparts. Interestingly, some Class~flat objects correlate well with high-temperature gas between $350-500$\arcsec separation, while many older sources appear in regions of lower gas temperature. 

However, $^{12}$CO$(3-2)$ is optically thick and predominantly traces warm gas near low-density H~\textsc{ii} regions. Therefore, we also look at the $(1-0)$ transitions of the commonly used dense gas tracers N$_2$H$^+$, HCN, and the more diffuse tracer HCN \citep{2017Pety}, for which the temperature can be linearly related to the gas density \citep{2023Tarfalla}. As seen in Fig.~\ref{fig:14-pv-diagrams}, the agreement between stars and gas components is generally good for HNC and HCN. Class~flat aligns best with regions of densest gas. While some Class~II and III YSOs are already quite dispersed from the gas velocities, others still show a tight correlation in radial velocity. The two Class~I objects are less in agreement with the gas velocities, but have significant error bars and low number statistics. 

There is little N$_2$H$^+$ emission, and the gas velocity seems slightly shifted compared to the velocities of the stars. These observations agree with analyses by \citep{2024Hacar, 2024A&ASocci}, who find that protostars in the Flame Nebula have a weaker positional correlation to N$_2$H$^+$ than other nearby regions, suggesting a rapid dispersal and photo-dissociation of dense gas by feedback from nearby massive stars like IRS 2b. As the ORION-B survey has a lower resolution than the ALCOHOLS survey, interpretations of the agreements are generally more difficult.

We find an overall good agreement between the gas and radial velocities of the young stars in our sample. While Class~III and some Class~II YSOs show a larger scatter, suggesting they may have started decoupling from the gas, Class~flat stars clearly maintain a connection to the gas in radial velocity. We can extrapolate that if the coupling in radial velocity remains intact for those YSOs, it should also be true for the transversal (proper) motions. This means that approximating the 3D motions of the Flame Nebula using, for instance, Class~II YSOs, of which many are visible in \emph{Gaia}, would be a reasonable assumption and should yield good estimates. Regarding the formation scenarios discussed in Sect.~\ref{sec:formation-comp}, the good agreement between gas and stellar velocities indicates that the theorized outward motion of older stars so far seems to be a mix between radial drift and dynamical heating, with Class~II and III objects likely transitioning towards the dynamical heating-dominated regime.

\section{Conclusions}
\label{sec:Conclusion}
In this study, we carried out the first homogeneous infrared proper motion calculation for stars in the direction of the Flame Nebula. Despite the challenge of deriving proper motions from ground-based data for sources at the cluster distance of $\sim420$ pc in the direction of the galactic anti-center, our results are of comparable quality to the \emph{Gaia} measurements for the region. This enabled us to perform the first homogeneous spatio-kinematic study of both its optically visible YSOs and the comparatively younger, still embedded stars.

We present an infrared-based proper motion catalog for 6769 stars in the direction of NGC~2024, among which we identified 362 YSO candidates. We confirmed the appearance of an age segregation and a mild gradient from young sources in the cluster center to older ones in the outskirts \citep{2014Getman}. 

Regarding the formation history of NGC~2024, our findings support an outward motion of older stars, driven partly by N-body interactions and consistent with the higher velocity dispersions of Class~II and III objects, but constrained by the fact that many YSOs seem to be still kinematically coupled to their surrounding gas. We find no evidence for inward motions associated with filamentary infall or remnants of hierarchical cluster assembly. All classes statistically share a common velocity distribution, with no trace of substructure detectable to date. This observation is consistent with a formation scenario involving either the rapid collapse of a string of subclusters into a monolithic structure or formation within a relatively coherent structure. The transversal velocity dispersions of Class~I YSOs are slightly higher than those of Class~flat, suggesting they may have formed during the rebound phase following the initial collapse.

We report a good agreement between the gas velocity and the radial velocities of our YSO sample. Some Class~II and III YSOs might be decoupling from the gas due to their larger scatter, while others retain a close connection. We extrapolate that, at least for the Flame Nebula, inferring 3D motions using proper motions of YSOs up to Class~II would be justified. 

With access to precise positional and velocity data of the youngest stars of a cluster, we are now at the precipice of resolving its star formation history for the first time. However, the surprising findings of a higher velocity dispersion of very young YSOs in L1688 and a hint of it in Orion highlight our knowledge gaps in the early stages of star formation. It also presents a pressing need for more radial velocity measurements of Class~I objects in NGC~2024, to be able to extrapolate the tentative trend seen in the proper motion dispersion towards the third velocity dimension and assess their coupling to the gas.

\section*{Data availability}
The full master catalog described in Sect.~\ref{sec:Results} will be available in electronic form at the CDS.

\begin{acknowledgements}
We thank the anonymous referee for the useful comments that helped to improve this publication.
Co-Funded by the European Union (ERC, ISM-FLOW, 101055318). Views and opinions expressed are, however, those of the author(s) only and do not necessarily reflect those of the European Union or the European Research Council Executive Agency. Neither the European Union nor the granting authority can be held responsible for them. AS acknowledges funding from the European Research Council (ERC) under the European Union’s Horizon 2020 research and innovation programme (Grant agreement No. 851435).
This work is based on observations collected at the European Southern Observatory under ESO programme(s) 60.A-9285, 179.A2010, and 198.C-2009.
This work has made use of data from the European Space Agency (ESA) mission \emph{Gaia} (\url{https://www.cosmos.esa.int/gaia}), processed by the \emph{Gaia} Data Processing and Analysis Consortium (DPAC, \url{https://www.cosmos.esa.int/web/gaia/dpac/consortium}).
\end{acknowledgements}
\bibliography{library}
\clearpage
\begin{appendix}
\section{Sample comparison and protostellar disk populations}
\label{appendix:Disks}

One of the earliest NIR surveys on the embedded center of NGC~2024, which remains an actively used member catalog, was conducted by \cite{1996Meyer}. Their original table of observed sources encompasses 233 objects, of which 216 were measured in at least one of the $JHK_S$ passbands. No membership indicators are included in this photometric observation list. Recently, using a spatially defined subset of 179 of the initially identified sources, \cite{Terwisga2020} observed protostellar disks in NGC~2024 with ALMA and presented evidence of two disk populations. They assumed each target was a disk-bearing source, but they only detected disks for 57 ($\sim 32\%$) of their targets. However, they inferred disk masses for detected disks, and mass limits using stacked observations for non-detections, and analyzed both under the assumption that all sources are disk-bearing. They presented two distinct disk populations, separated into a more eastern population with a higher detection rate of 45\% and generally more massive disks, and a second smaller population of less massive disks with a detection rate of only $\sim 15\%$ towards the West of the nebula.

Given the continued usage of the \cite{1996Meyer} catalog, we compare it to our observations and determine how many objects in their list are classified as YSOs in the \citep{2025Roquette} literature catalog. Using $r=2$\arcsec, we crossmatch 146 objects from the input catalog, including seven duplicated sources (seven pairs; identical or almost identical position coordinates). Of the uniquely matched sources, 89 are in the \cite{2025Roquette} catalog (78 bona fide), and 16 have \emph{Gaia} measurements. Almost half of the bona fide matches (34) are Class~I, followed by 22 Class~flat, 19 Class~II, and only three Class~III (thin-disk) objects. This means that we can only assign a YSO class to ca. 35\% of the objects of \cite{1996Meyer} as YSOs from today's point of view.

For 71 stars with photometric measurements, we report no crossmatch. An investigation of the available magnitude values indicates that around 50 could plausibly fall within the brightness range our study is sensitive to. A possible explanation of why they are missing in the catalog could be, for example, the characteristic variability in the brightness of young stars, which could prevent them from being detected in at least 9 of our epochs, or the positional accuracy of the \cite{1996Meyer} measurements to within 2 arcseconds.

Next, we compare our catalog with the catalog of detected and undetected disks published by \cite{Terwisga2020}: A crossmatch ($r=2$\arcsec) between our catalog and theirs yields an overlap between 69 of 122 undetected disks and 44 of 58 detected disks. It should be noted that 13 of the sources that are both detected disks and in our catalog are not marked as YSOs by \citep{2025Roquette}. This points to a possible misclassification based on the IR excess. Of the undetected disks, 24 are not classified as YSOs, but since no disks could be detected in the ALMA observations, a misclassification in their case is not as likely as for the detected disks. In total, our NIR catalog encompasses $113/180 \approx 63\%$ of theirs. Of the non-matches, 16 are listed as YSO candidates in the NEMESIS catalog, but are missing from our sample. 

\begin{figure}[ht]
    \centering
    \includegraphics[width=\linewidth]{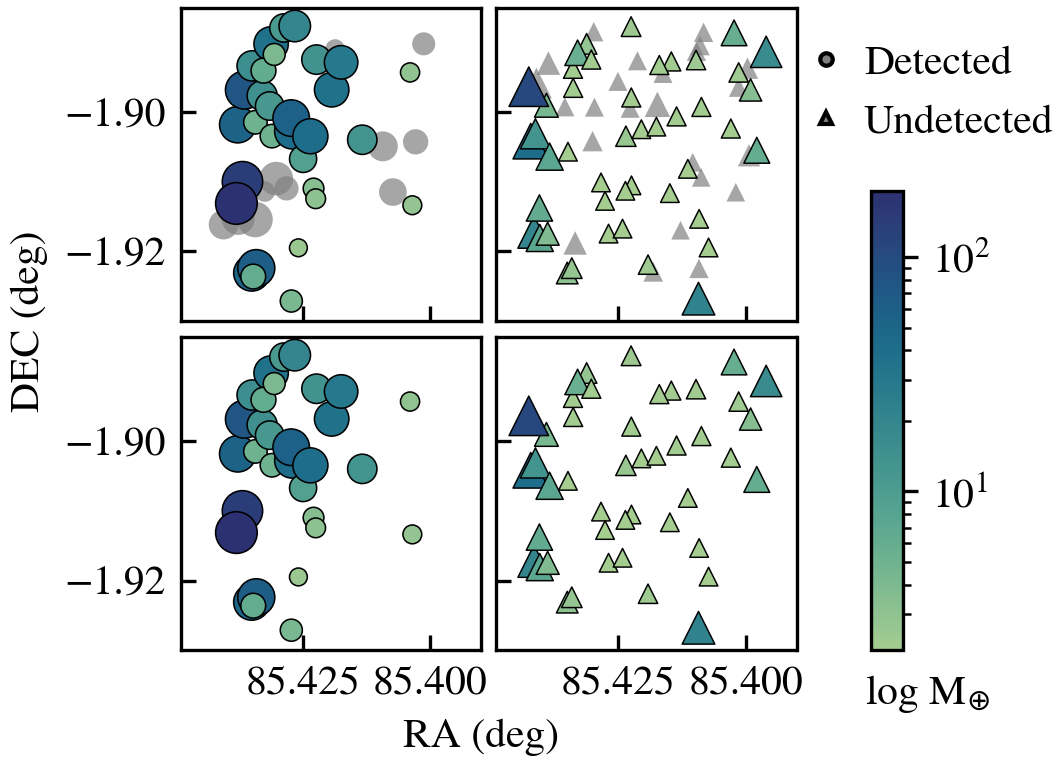}
    \caption{Positions and estimated disk masses for the subset 144 of disk-bearing stars identified by \citep{Terwisga2020} that are also in our catalog. The sources that are not classified as YSO candidates in \cite{2025Roquette} are indicated in gray in the top row and omitted in the bottom row. The source sizes and colors correspond to their estimated masses, in units of Earth masses. The left column shows sources for which disks were detected with ALMA, while the right plot displays stars with no detected disks, where a minimum mass was derived from flux measurements.}
    \label{fig:08-YSO-diskpop}
\end{figure}

Fig.~\ref{fig:08-YSO-diskpop} shows the positions of the crossmatched sample, color-coded and sized according to the respective disk masses, using a logarithmic scale. Circles in the left panel represent the YSOs where disks were detected with ALMA, while triangles in the right plot denote the sample where no disks were detected. This display aims to assess whether the previously described results of \cite{Terwisga2020} of two disk populations can be reproduced using only our bona fide YSO sample.

We concur that disk masses are generally higher towards the East of the Nebula. This is not surprising as the dense molecular ridge of the Nebula is located there \citep{2008Meyer, Terwisga2020}. However, there are also some high mass limits for the sample of non-detections in the West, while a lot of YSOs with non-detected disks and small mass limit estimates fill the space between the two extremes. We show that the previously reported mass trend also appears in our reduced sample of $\sim 60\%$ of the original sample size. We can not tell whether it truly indicates the presence of two distinct disk populations or simply because more dense molecular gas is available towards the East of the Flame Nebula. When only looking at the detections versus the non-detections, using the bona fide sources, we find that the median proper motion for the first group $(0.39,-0.89)$ mas\,yr$^{-1}$, and for the second it is slightly different, at $(-0.14, -1.13)$ mas\,yr$^{-1}$. A more in-depth study is needed to determine whether this shift is statistically significant and corresponds to an actual physical effect.

\clearpage
\newpage
\section{Spatial dispersion, internal motion and size-velocity relation of the NGC~2024 YSOs}
\label{appendix:internal-motion}

\subsection{Spatial dispersion}

Tab.~\ref{tab:YSO-covariances} lists the spatial dispersions and 95\% confidence ellipse parameters for the four YSO classes shown in Fig.~\ref{fig:09-YSO_classes_Gaia-kde}. The 1$\sigma$ uncertainties in $\alpha$ and $\delta$, along with their covariance, quantify the spread of each class, while the semi-major ($a_{95}$) and semi-minor ($b_{95}$) axes describe the overall 95\% confidence ellipse. Younger YSOs are more tightly clustered, whereas older YSOs show larger dispersions and more elongated distributions, indicating progressive dispersal from their birth sites consistent with cluster dynamical evolution.

\begin{table}[h!]
\centering
\caption{Spatial dispersions and 95\% confidence ellipse parameters for the YSO classes. The covariance is given in arcmin$^2$, all other values in arcmin.}
\begin{tabular}{lcccccc}
\hline \hline
Class & $\sigma_{\rm \alpha}$ & $\sigma_{\rm \delta}$ & ${\rm cov(\alpha,\delta)}$ & $a_{95}$ & $b_{95}$ & $\sigma_{\rm \delta}$ / $\sigma_{\rm \alpha}$\\
\hline
I & 4.24 & 2.78 & -0.03 & 10.38 & ~6.80 & 0.66\\
flat & 3.85 & 4.77 & ~2.33 & 11.84 & ~9.22 & 1.24 \\
II & 4.78 & 6.03 & ~0.32 & 14.75 & 11.70 & 1.26\\
III & 7.57 & 8.16 & 11.60 & 21.11 & 17.21 & 1.08 \\
\hline
\end{tabular}
\label{tab:YSO-covariances}
\end{table}

\begin{figure*}[ht]
    \centering
    \includegraphics{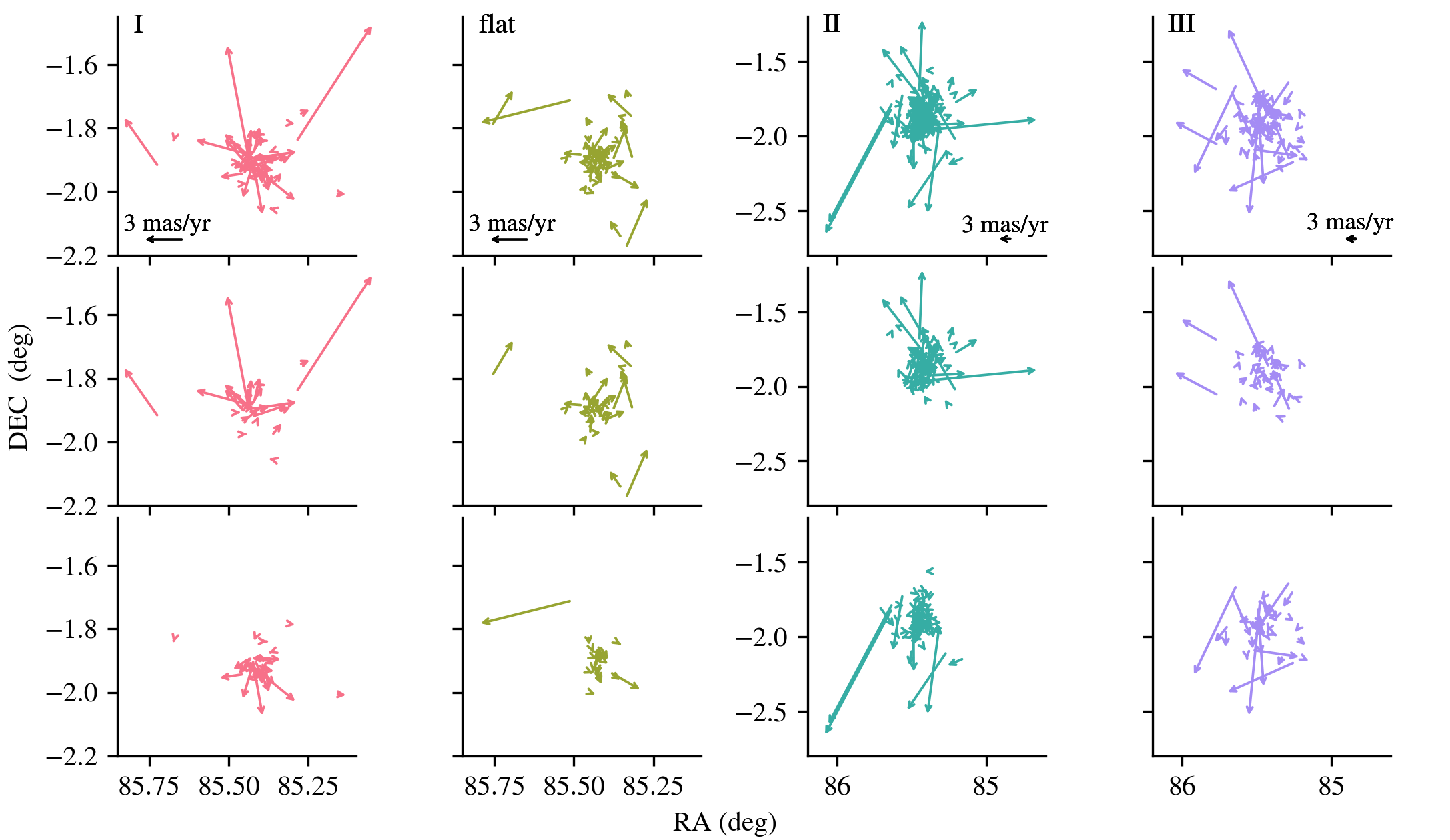}
    \caption{Position-vector diagrams of the internal motions of the NGC~2024 YSOs, divided by YSO class. For better visual differentiation, the second row shows only sources with positive motion in declination, and the third row shows stars with negative motion in declination. Note that the axis scale differs between Class~I/flat and Class~II/III objects.}
    \label{fig:09-YSO_vector_plot_LSR}
\end{figure*}

\subsection{Internal cluster motion}

To analyze the internal cluster motions and relate them to the stellar positions, we transformed the proper motions we calculated to the Local Standard of Rest (LSR), assuming a cluster distance of $d\sim 420$ pc of the cluster \citep{2019Zucker} for stars without \emph{Gaia} parallax measurements. We subtracted the mean proper motion of all 362 bona fide members from each proper motion measurement. The bulk-motion corrected proper motions in the LSR are shown as vectors in Fig.~\ref{fig:09-YSO_vector_plot_LSR}. The panels for Class~I/flat sources are zoomed in due to the more centralized positions of those stars. 

Except for Class~III sources, most stars expand roughly isotropically from a central region. Class~I and flat-spectrum objects show smaller proper motions than Class~II sources, while Class~III stars are more dispersed and no longer exhibit a clear expansion pattern. Class~II sources are hard to analyze due to the crowding of many sources with small proper motions in the center. Still, the stars with larger proper motions preferentially move along the N-S axis, corresponding to the ``wick'' or central dark lane of the Flame Nebula. This could mean that the presence of the gas reservoir along this axis influenced the Class~II motions, but the younger YSO generations no longer feel the effect.

The spatio-kinematic distribution of the YSO classes shows no clear evidence for subclusters or their remnants within our sample.

We note that three Class~I stars and one flat-spectrum object exhibit unusually high proper motions for their young ages compared to the rest of their respective populations. These sources could either be runaway star candidates or contaminants in our sample. A more detailed investigation could be the subject of a future follow-up study.

\subsection{Velocity-size relation}
\label{appendix-Larson}

Pioneering work by \citet{1981Larson} on the physical properties of molecular clouds showed that they obey three empirical scaling relations involving their characteristic size $L$ (pc), mass $M$ (M$_\odot$), velocity dispersion $\sigma_v$ (km\,s$^{-1}$), and density $n$ (cm$^{-3}$):
\begin{align} 
\sigma_v &\propto L^{0.38}, \label{rel:1}\\
\sigma_v &\propto M^{0.2},\\
n &\propto L^{-1.1} \label{rel:3} 
\end{align}
These relations have been repeatedly confirmed in subsequent studies, both within the Milky Way and in external galaxies \citep[e.g.,][]{1987Solomon, 2008Bolatto, 2009Heyer}, although with minor variations. For instance, the exponent in Eq.~\ref{rel:1}, commonly denoted as $\alpha$, was found to be steeper ($\alpha \approx 0.5$) in later work \citep{1987Solomon}. However, more recent studies using stars as proxies for molecular clouds in the Scorpius-Centaurus OB association report an even steeper relation ($\alpha = 0.66$) and highlight the difficulties in interpreting Eq.~\ref{rel:1}, given the heterogeneous methods used to derive $\alpha$ throughout the literature \citep{2025Grossschedl}.

Because star formation is intrinsically connected to molecular clouds, and because we are studying very young objects, it is reasonable to investigate whether the scaling relations attributed to molecular clouds are also reflected in our YSO sample. Since our analysis is based on stellar astrometry, we focus on the velocity–size relation (Eq.~\ref{rel:1}).

Our measurements provide only first-order approximations, as we use the two-dimensional velocity dispersion of the YSOs, $\sigma_{v, \text{2D}}$, instead of the commonly assumed three-dimensional velocity dispersion. We choose this approach because otherwise the sample would be strongly biased against the youngest objects, for which radial velocities are available only for two Class~I and seven Class~flat objects. Furthermore, as noted in \cite{2025Grossschedl}, different methods for computing the cumulative velocity dispersions and cloud sizes in the literature can affect the derived $\alpha$ values. As described in the main text, we compute the two-dimensional velocity dispersion as the square root of the trace of the velocity covariance matrix $\sigma_{v, 2D} = \sqrt{(\sigma_{v,\alpha^*})^2+(\sigma_{v, \delta})^2}$. To mitigate methodological effects in the cluster size estimates, we approximate the cluster size (1) using the root-mean-square (rms) radius $r_{\text{rms}}$ assuming spherical geometry, and (2) using the effective radius $r_{\text{area}}$ of the convex hull enclosing the stellar positions.  

\begin{figure}[ht]
    \centering
    \includegraphics[width=\linewidth]{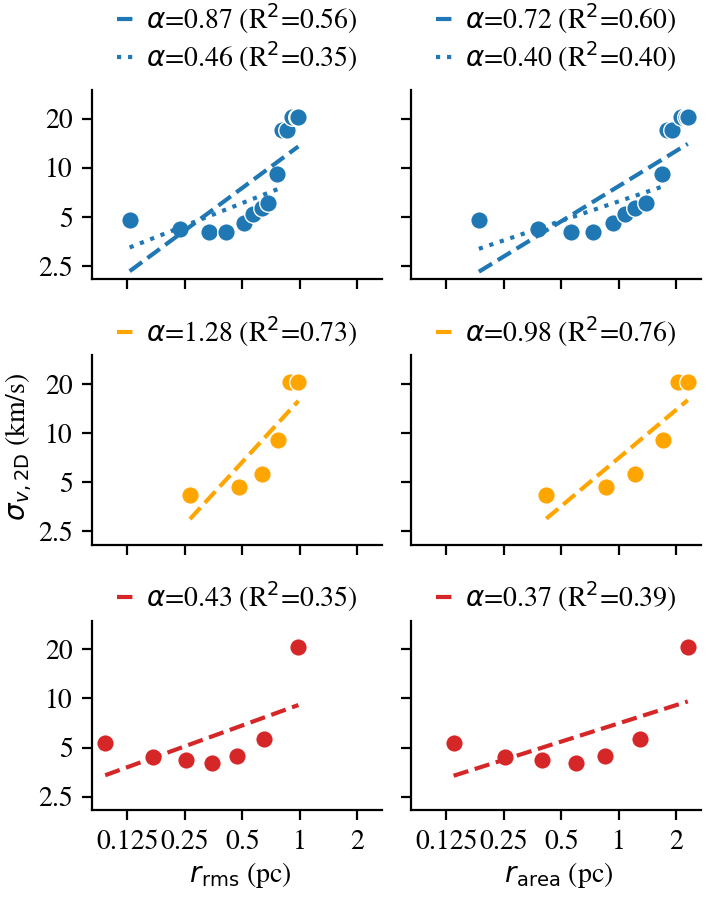}
    \caption{Two-dimensional velocity–size distributions of the YSO candidates for three binning strategies, shown with the associated power-law fits (dashed line).
Top: 15 linear radial bins, with an additional fit to the inner 10 bins (dotted line) to test sensitivity to outer outliers.
Middle: 7 linear radial bins.
Bottom: 7 bins with equal star counts.}
    \label{fig:Larson}
\end{figure}

We grouped all YSOs into cumulative radial bins centered on the two-dimensional cluster center. We tested two definitions of the cluster center, $(85.429,-1.912)$\footnote{\url{https://www.eso.org/public/images/eso0949n/}} and the mean of the YSO coordinates $(85.432, -1.893)$, which yielded consistent results. Three binning strategies were employed: two linearly spaced schemes with 15 and seven bins, and one equal-number binning scheme with seven bins. The results are shown in Fig. \ref{fig:Larson} for both radius calculations. 

The plots show that the data do not support the expected power-law relation, as indicated by consistently low $R^2$ values across all binning strategies. The $R^2$ indicates how much of the variance in the data is explained by the fit, with $R^2=1$ corresponding to a perfect fit.
The quality of the fit degrades substantially when using more than seven bins or when adopting the equal-number binning scheme. Although the latter strategy formally yields $\alpha$ values in the canonical range of $0.38-0.5$, these values are unreliable because the power-law fit explains little of the variance in the measured velocity dispersions. For the linearly spaced bin fits for the full range (dashed lines), the fits have higher $R^2$ values, but the resulting $\alpha$ values are generally higher than the canonical range. The area-based radius estimate produces slightly lower $\alpha$ values than $r_{\text{rms}}$, but the $R^2$ values, i.e., the fit quality, are nearly identical between the two radius estimates.

We also considered that velocity dispersions in the outer regions could be influenced by outliers. Therefore, we repeated the fit using only the innermost 10 of the 15 linearly spaced bins (dotted line) in the top row of Fig.~\ref{fig:Larson}. However, the derived $\alpha$ values remain unreliable because of the poor quality of the fit. Additionally, the slopes and fit quality are highly sensitive to the choice of center coordinates; however, better fits $(R^2> 0.8)$ were only obtained for implausible center positions. Depending on the binning strategy and center coordinate choice, these fits yielded inconsistent $\alpha$ values, ranging from 0.5 to 1.43. Performing the same analysis per evolutionary class likewise reveals no evidence for a consistent scaling relation analogous to Eq.~\ref{rel:1}.

We therefore conclude that our stellar sample does not follow the velocity-size relation $\sigma_v \propto L^{\alpha}$, with $\alpha = 0.38-0.5$. The NGC~2024 cluster and the Flame Nebula are relatively small systems, and similar deviations from the classical Larson relations have already been reported on such scales. In particular, \cite{2010Lombardi} showed that Eq.~\ref{rel:3} does not hold for individual cores and molecular clouds because their column densities cannot be described as constant. Since $n$ and $\sigma_v$ are directly linked through Eqs.~\ref{rel:1} and \ref{rel:3}, a breakdown of the velocity-size relation on these small scales is therefore expected, consistent with our findings.

\clearpage
\newpage
\section{Instrumentation and observing parameters}
\label{appendix:instrumentation-observation}

\begin{table*}[ht]
    \caption{Observing parameters for the 11 datasets.}
    \centering
\begin{tabular*}{\linewidth}{@{\extracolsep{\fill}}cccccccccc}
\hline \hline
Obs. Date    & Filter    & DIT   & NDIT  & NJITTER   & Single exp.   & $t_{\mathrm{exp}}$    &  N\tablefootmark{a}    & Image Quality\tablefootmark{b}    & Mean Seeing   \\
(dd/mm/yyyy) &           & (s)   & (\#)  & (\#)      & time (s)      & time (s)          & (\#)  & (arcsec)                                      & (arcsec)      \\
 \hline
 \multicolumn{10}{c}{\emph{Science Verification galactic mini-survey (SV)}} \\
 \hline
20/10/2009        & J         & 4     & 8     & 2         & 32            & 128               & 12            & 0.853                             & 0.848         \\
24/10/2009        & H         & 2     & 12    & 2         & 24            & 96                & 12            & 0.926                             & 1.228         \\
20/10/2009        & Ks        & 2     & 12    & 2         & 24            & 96                & 12            & 0.783                             & 0.657         \\
 \hline
 \multicolumn{10}{c}{\emph{Vista Hemisphere Survey (VHS)}} \\
 \hline
24/11/2014        & J         & 15    & 1     & 2         & 15            & 60                & 12            & 0.712                             & 0.573         \\
24/11/2014        & Ks        & 7.5   & 2     & 2         & 15            & 60                & 12            & 0.691                             & 0.515         \\   
 \hline
 \multicolumn{10}{c}{\emph{VISIONS}} \\
 \hline
13/11/2017        & H         & 3     & 2     & 5         & 6             & 60                & 30            & 0.835                             & 0.568         \\
17/02/2018        & H         & 3     & 2     & 5         & 6             & 60                & 30            & 0.969                             & 0.996         \\
12/10/2018        & H         & 3     & 2     & 5         & 6             & 60                & 30            & 1.247                             & 1.040         \\
17/02/2019        & H         & 3     & 2     & 5         & 6             & 60                & 30            & 0.633                             & 0.531         \\
08/11/2019        & H         & 3     & 2     & 5         & 6             & 60                & 27\tablefootmark{c}            & 0.846                             & 0.521         \\
16/02/2020        & H         & 3     & 2     & 5         & 6             & 60                & 30            & 0.769                             & 0.479         \\
\hline
\end{tabular*}
\tablefoot{
\tablefoottext{a}{Number of individual images used to create a tile: N = number of pawprints (6) $\times$ NJITTER.}\\
\tablefoottext{b}{Computed by multiplying the pixel scale with the mean FWHM determined via \texttt{IRAF/imexam}.}\\
\tablefoottext{c}{Three exposures could not be used due to bad quality and high readout noise of some detectors.}}
\label{tab:02-Observing-parameters}
\end{table*}

\subsection{Instrumentation}

All data were collected with the 4.1m Visible and Infrared Survey Telescope for Astronomy (VISTA) on the ``NTT peak'' at Cerro Paranal in Chile \citep{2006Emerson}. It hosted the VISTA InfraRed CAMera (VIRCAM) \citep{2006Dalton}, a wide-field near-infrared camera facility ($0.8-2.3$ $\mu$m) mounted on the telescope from 2008 until its decommissioning in early 2022. The camera consisted of a sparse 4$\times$4 matrix of 16 detectors with a pixel size of $\sim 0.34$ arcseconds. The detectors were characterized by a $1-10$\% deviation from linearity and saturation limit between 24,000 and 37,000 ADU \citep{2010Arnaboldi}. The camera had a seeing-limited spatial resolution, and the best achievable image quality after considering the effects of seeing, optics, and sampling was around 0.6 arcseconds. Its $JHK_s$ passbands are characterized as follows: $J$ ($\lambda_c = 1.25~\mu$m, FWHM = 0.18 $\mu$m), $H$ ($\lambda_c = 1.65~\mu$m, FWHM = 0.3 $\mu$m), and $K_S$ ($\lambda_c = 2.15~\mu$m, FWHM = 0.3 $\mu$m).

Further specifications of the observing instrument can be found in \cite{2004Emerson, 2006Dalton, 2006Emerson}, on the ESO website\footnote{\url{https://www.eso.org/public/teles-instr/paranal-observatory/surveytelescopes/vista/camera/}\\\url{https://www.eso.org/sci/facilities/paranal/decommissioned/vircam/inst.html}}, or in the VIRCAM user manuals\footnote{\url{https://www.eso.org/sci/facilities/paranal/decommissioned/vircam/doc.html}}.

\subsection{Observing scheme}

An observation with VIRCAM worked as follows: To create a contiguous $\sim1.5^{\circ} \times 1.0^{\circ}$ ``tile'' (see Fig.~\ref{fig:01-FN-coverage-tiles}), the smallest regular observational field, a stack of six individual images, called ``pawprints'', was captured. The pawprints were taken with pre-determined offsets, ensuring that the majority of each sky position within a tile was imaged at least twice. As detailed in \cite{2004Emerson}, a small, unspecified margin at the Y-axis limits of a tile remains covered only once in six pawprints. Furthermore, a pattern of smaller offsets, called jitters, was executed at each pawprint position to support the rejection of bad pixels during the co-addition process of a tile. Thus, the specific coverage of different sky positions varies across the tile \citep[see e.g.,][ Fig.~4]{Meingast2023b}. 

The typical effective exposure time across the tile is $t_{\text{exp}}~\text{(s)} = 2\times\text{DIT}\times\text{NDIT}\times\text{NJITTER}$, where NDIT refers to the number of single detector integration times (DITs) that are added for one readout of the detector, and NJITTER indicates the number of jittered positions at each pawprint location.

\subsection{Tables of observing parameters and saturation limits}

The observing parameters for the 11 datasets used in this study are listed in Tab.~\ref{tab:02-Observing-parameters}. The saturation limits used for source list cleaning (Fig.~\ref{fig:02-Workflow-dataset}) are listed in Tab.~\ref{tab:03-Saturation-limits}. For the PSF-fitting method, the saturation limit of each dataset was determined via \texttt{imexam}. For the \texttt{vircampype} method, they were determined during the routine \citep{Meingast2023b}. 

\begin{table}[ht]
    \caption{Saturation magnitude thresholds for the 11 datasets for sources detected with \texttt{vircampype} (centroiding), and PSF-fitting, respectively.}
    \centering
    \begin{tabular}{lcc}
        \hline\hline
        Dataset & \multicolumn{2}{c}{Saturation limits (mag)} \\
         & \texttt{vircampype} & PSF \\
        \hline
        SV $J$      & 12.0 & 11.8 \\
        SV $H$      & 11.5 & 11.3 \\
        SV $K_S$    & 11.0 & 10.7 \\
        VHS $J$     & 13.0 & 13.7 \\
        VHS $K_S$   & 12.0 & 13.0 \\
        VISIONS A   & 12.0 & 12.8 \\
        VISIONS B   & 12.0 & 12.3 \\
        VISIONS C   & 12.0 & 12.1 \\
        VISIONS D   & 12.0 & 12.5 \\
        VISIONS E   & 12.0 & 12.5 \\
        VISIONS F   & 12.0 & 12.3 \\
        \hline
    \end{tabular}
    \label{tab:03-Saturation-limits}
\end{table}

\clearpage
\newpage
\section{IRAF pipeline}
\label{appendix:IRAF}

The \texttt{vircampype} astrometric pipeline was specifically developed to meet the requirements of the VISIONS survey \citep{Meingast2023b}. It outperformed the CASU pipeline in various tests, but its efficiency in crowded environments, such as star clusters, which could impede the centroiding procedure used for position determinations, has not previously been tested.

To benchmark pipeline performance in these environments, we carried out an independent astrometric reduction of the Tile$^*$ images for all 11 datasets using PSF-fitting. This produced a second source catalog for each region, which we then compared to the corresponding \texttt{vircampype} catalog.

We used the \texttt{daophot} routine implemented in the large, general-purpose astronomical software Image Reduction and Analysis Facility (\texttt{IRAF}) Version 2.17\footnote{\url{https://iraf.readthedocs.io/en/latest/releases/v217revs.html}}. We note that this implementation is freely available as a \texttt{docker} image\footnote{\url{ https://hub.docker.com/r/smeingast/iraf}}.

\subsection{Source catalog generation}

The extraction of stellar positions happened in six steps, each corresponding to a \texttt{daophot} subroutine:

\begin{enumerate}
    \item Image parameter determination --- \texttt{imexam}
    \item Source selection --- \texttt{daofind}
    \item Aperture photometry --- \texttt{phot}
    \item Selection of stars for PSF model --- \texttt{pstselect}
    \item Creation of a model PSF --- \texttt{psf}
    \item Positions and magnitudes for all sources --- \texttt{allstar}
\end{enumerate}

We first measured the stars' typical full-width half-maximum (FWHM), the mean background deviation, and the minimum and maximum pixel values with \texttt{imexam}. Next, we performed the \texttt{daofind} routine with a $5\sigma$ peak detection threshold and a center box $\texttt{cbox}=2\times \textrm{FWHM}$ to generate a source list. We performed aperture photometry on the extracted sources using the \texttt{phot} routine with five different aperture sizes ($1-5 \times$ FWHM), and sky annulus sizes according to the recommendation in the IRAF user manual for CCD photometry\footnote{\url{https://iraf.readthedocs.io/en/latest/extradocs.html}}. Next, we selected candidate stars for the PSF model with the \texttt{pstselect} routine. This routine automatically provides the chosen number $n$ brightest stars in the source list. However, this also includes saturated stars and is not wholly representative of the stellar content in the image. Therefore, we set $n=4000$ and randomly sampled $100$ sources from the provided list. We used the randomly sampled stars as input for the \texttt{psf} routine, adapting the PSF radius and fitting scale radius from the user manual. We inspected the surface plots of the 100 samples and retained them if the peak was well-defined and isolated, or rejected them if not. Due to the excellent seeing conditions, many observation epochs were nearly undersampled with very narrow FWHM values. Thus, we allowed for a slight asymmetry in the shapes of the peaks. A model PSF was built from fitting the retained sources, typically between 40 and 60\% of the initial sample, with an analytical function. We used all five pre-defined analytical functions implemented in \texttt{IRAF/daophot} to generate model PSFs, which were subtracted from a given tile using the \texttt{allstar} command. The performance of the various models was compared based on their norm scatter values, and the best fit was used. Lastly, the subtracted image was inspected visually to determine the quality of the astrometric results.

\subsection{Spurious source removal}

Before running \texttt{daofind}, we defined a saturation threshold for each image using \texttt{imexam}. This ensured that the centers of the saturated sources were not in our source list, but we discovered many spurious detections appearing in their halos, likely caused by bleeding across pixels. Unlike other kinds of spurious sources, such as variations in the sky background or remnants of nebulosity, these measurements were not always removed by the filter criterion $N_{\text{occ}} \geq 2$. To remove these misdetections, we inferred the positions of all sources brighter than the saturation limits (Tab.~\ref{tab:03-Saturation-limits}) using the \texttt{vircampype} source catalog, in which bright sources are replaced with \emph{2MASS} measurements. We then empirically defined a magnitude-dependent radius around each of the \emph{2MASS} sources in our field of view brighter than $14$ mag and removed multiple source detections within this area.

\subsection{Positional uncertainties}

The \texttt{allstar} routine does not provide error estimates for the extracted stellar positions, which would lead to significantly underestimated errors in proper motion for the PSF-fitted sources. So we expanded on a concept described in \cite{2009Andersen, 2017Andersen}, where the authors performed completeness corrections and estimated statistical photometric errors by adding clones of their sources to their images, and re-extracting them.

\begin{figure}[h]
    \centering
    \resizebox{\hsize}{!}{\includegraphics{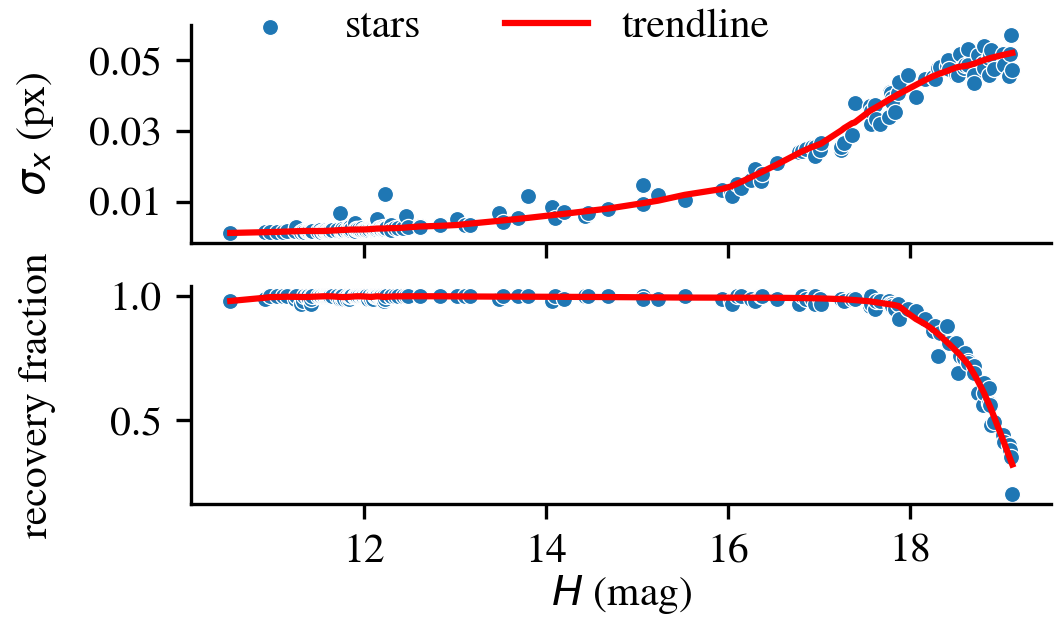}}
    \caption{Statistical position error in the $x$ pixel coordinate (top) and recovery fraction (bottom) as a function of source brightness..}
    \label{fig:recovery-example}
\end{figure}

Similarly, we randomly split our sample into seven subsets and added $n=100$ clones of each star in a subgroup to a Tile$^*$ at random locations using \texttt{addstar}. The subsets ensured that we added no more than 30\% of the original source list size to the image, thereby artificially inducing crowding. We then performed the \texttt{daofind, phot} and \texttt{allstar} routines again. We calculated the statistical measurement uncertainty $e_{\text{stat}}$ from the standard deviation between input and output locations of the clones. We could also determine a recovery fraction for each star from this procedure. An example of $e_{\text{stat},x}$ and recovery fraction as functions of source brightness is given in Fig.~\ref{fig:recovery-example} for the SV $H$-band.

This uncertainty cannot account for all systematic effects and error contributions outside the method. However, these effects are addressed in the \texttt{vircampype} workflow. We added a systematic error floor $e_{\text{sys}}$ to each statistical uncertainty to get a better overall error estimate on the positions. Its value was determined as the minimum positional errors in $\alpha$ and $\delta$ for each dataset using the \texttt{vircampype} source catalog. The final errors were computed as 
$e_{\text{pos}}= \sqrt{e_{\text{sys}}^2 + e_{\text{stat}}^2}$. The same procedure was used to calculate the magnitude errors for PSF sources.

\clearpage
\newpage
\section{Bayesian inference model}
\label{appendix:Bayesian-model}

\begin{figure*}[t!]
    \centering
    \includegraphics{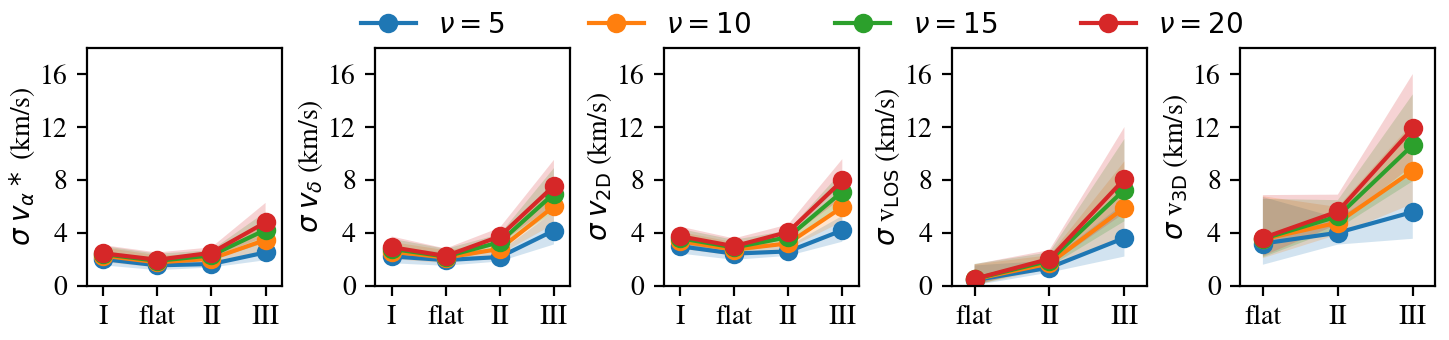}
    \caption{Evolution of the velocity dispersions of the different YSO classes, assuming a Student-t likelihood in the Bayesian inference model. The colored markers and shaded areas indicate the posterior median velocity dispersion, along with the 95\% credible intervals for various settings of the degrees of freedom parameter $\nu$. Insufficient Class~I data are available for the two rightmost panels.}
    \label{fig:14-velocity-params-tdist}
\end{figure*}

We aim to model the underlying three-dimensional velocity distribution of the stars in our bona fide YSO sample to determine the intrinsic velocity dispersion of each evolutionary class. Each class contains a limited number of stars: $54-171$ for tangential and 2D velocities, and $7 - 42$ for radial velocities and the full 3D velocity sample, due to the limited available radial velocities \citep{2025Roquette}. A Bayesian framework is ideal for this analysis, as it allows for a principled treatment of uncertainty. The velocity distribution of young stellar populations can be described by a multivariate normal distribution $\mathcal{N}$, with the covariance matrix $\mathbf{S}$ representing the intrinsic dispersion. The observed velocities can then be modeled as the convolution of this intrinsic normal distribution with the observational heteroscedastic error distribution, meaning each source has its own measurement error $\mathbf{C}_i$. The resulting likelihood can be written as $\mathbf{v}_i \sim \mathcal{N}\!\left(\boldsymbol{\mu},\, \mathbf{S} + \mathbf{C}_i\right)$.

Our goal is to infer the unknown mean and covariance matrix of the error-deconvolved velocity distribution for each dimension. The posterior space includes up to three dimensions from the mean and six unique dimensions from the covariance matrix. Since the covariance matrix must be symmetric and positive semi-definite, we use the Cholesky decomposition for numerical stability. By parameterizing the covariance matrix $\mathbf{\Sigma}$ via its lower triangular Cholesky matrix $L$, we ensure positive definiteness and allow for correlations among the velocity components. For each dimension $d = 1, \dots, 3$, we construct $\mathbf{L}$ from the parameters $\{l_{ij}, r_{ij}\}$ such that $\mathbf{\Sigma} = \mathbf{L} \mathbf{L}^{\mathsf{T}}$. This transformation maps real-valued parameters to valid covariance and correlation structures. We use a multivariate normal distribution as our primary likelihood function, but also test a multivariate Student-t likelihood to assess the robustness of the results against possible outliers.

We assign Gaussian priors to the model parameters as follows: For each velocity component $i = 1, \dots, d$, the location parameters have priors $\mu_i \sim \mathcal{N}(0, 10^2)$, and the diagonal elements of the Cholesky factor are defined as $l_{ii} \sim \mathcal{N}(0, 2^2)$. For dimensions $d > 1$, the off-diagonal correlation parameters $r_{ij}$ ($i < j$) are assigned Fisher-z priors $r_{ij} \sim \mathcal{N}(0, 1.5^2)$. The Fisher-z transform is a monotonic transformation that maps a correlation coefficient $r$, which is bounded in $[-1, 1]$, to $(-\infty, \infty)$, making inference numerically stable. The prior is defined on the transformed space, effectively covering the full back-transformed $r$ interval.

Posterior sampling is carried out using the No-U-Turn Sampler (NUTS) \citep{10.5555/2627435.2638586} implemented in \texttt{blackjax} \citep{2024blackjax}, with a warm-up phase to adapt the step size and the number of leapfrog steps. The sampler then generates 10,000 samples from the posterior distribution. For each sample, the velocity dispersion is calculated as the square root of the trace of the modeled covariance matrix. The final velocity dispersion for each parameter and YSO class is taken as the posterior median, with a 95\% credible interval determined from the 2.5$^{\text{th}}$ and 97.5$^{\text{th}}$ percentiles.

The results using the multivariate normal likelihood are shown in Fig.~\ref{fig:10-velocity-params} in the main text. However, due to the high velocity dispersions of Class~III objects in this model, we tested the robustness of the results derived using the Gaussian model against potential outliers by also modeling the underlying velocity distribution with a multivariate Student-t likelihood. The Student-t distribution introduces an additional parameter, the degrees of freedom $\nu$, which controls the heaviness of the distribution's tails. As $\nu \rightarrow \infty$, it converges to a Gaussian distribution. As $\nu$ decreases, the model accommodates more points as outliers, effectively increasing the weight of central data points. We repeated the inference procedure over a range of $\nu \in [1, 30]$ and show the results for four fixed values in Fig.~\ref{fig:14-velocity-params-tdist}.

For all $\nu > d$, where $d$ is the dimension of the investigated velocity parameter, respectively, we qualitatively recover the same trends as for the Gaussian model. The inferred velocity dispersions become slightly smaller if more outliers are accommodated, but the overall kinematic trends remain consistent across different YSO classes and velocity parameters. In particular, Class~I sources consistently have higher velocity dispersions than Class~flat  sources for all $\nu$ values. However, the significance of the difference between the two values decreases for smaller $\nu$. Nevertheless, the dispersions are at least comparable, deviating from the general trend of increasing dispersion with more evolved YSO classes between Class~flat and Class~III. For Class~II and Class~III, increasing $\nu$ slightly increases the reported velocity dispersions. In 1D, the velocity dispersion of Class~II is even lower than that of Class~I for $\nu < 8$ $(\sigma_{v, \delta})$ and $\nu < 20$ $(\sigma_{v, \alpha^*})$. Overall, Class~I and Class~II velocity dispersions are generally comparable in 1D and 2D, when the model allows for more outliers.

Because the presence of velocity outliers in the dataset is uncertain, and likely minimal for the Class~I and Class~flat populations, which are located in the central nebulosity of the Flame Nebula, we cannot definitely rank Class~I and Class~II velocity dispersions.
Nevertheless, the consistency of inferred dispersion trends across the full $\nu$ range with the trends shown in Fig.~\ref{fig:10-velocity-params} demonstrates that our results are robust against deviations from Gaussian velocity distributions.

\end{appendix}
\end{document}